\documentclass[aps,preprint,superscriptaddress,groupedaddress]{revtex4}  % for review and submission

\usepackage{graphicx}  % needed for figures
\usepackage{dcolumn}   % needed for some tables
\usepackage{bm}        % for math
\usepackage{amssymb}   % for math
\usepackage{epstopdf}
\usepackage{subfig}
\usepackage{xspace}
\usepackage{amsmath}
\usepackage{color}

\newcommand{\fastjet}{\textsc{FastJet}\xspace}

\newcommand{\zg}{\ensuremath{z_G}\xspace}
\newcommand{\rg}{\ensuremath{R_G}\xspace}
\newcommand{\Rg}{\ensuremath{R_G}\xspace}
\newcommand{\ME}{\ensuremath{M/E}\xspace}
\newcommand{\MgE}{\ensuremath{M_G/E}\xspace}

\newcommand{\sherpa}{\textsc{Sherpa}\xspace}
\newcommand{\herwig}{\textsc{Herwig}\xspace}
\newcommand{\pythia}[1]{\textsc{Pythia}{#1}\xspace}
\newcommand{\pyquen}{\textsc{Pyquen}\xspace}

\begin{document}

% The following information is for internal review, please remove them for submission
\widetext
\leftline{Version 3 as of \today}
\leftline{Primary authors: MIT, NTU, INFN Bari, GSU, CU Boulder, CERN}
%\leftline{Publication: TBD}
\leftline{MITHIG-MOD-21-001}

\title{Jet energy spectrum and substructure in $e^+e^-$ collisions at 91.2 GeV with ALEPH Archived Data}

\author{Yi Chen}
\email{chenyi@mit.edu}
\affiliation{Massachusetts Institute of Technology, Cambridge, Massachusetts, USA}%

\author{Anthony Badea}
\affiliation{Harvard University, Cambridge, Massachusetts, USA}%

\author{Austin Baty}
\affiliation{Rice University, Houston, Texas, USA}%

\author{Paoti Chang}
\affiliation{National Taiwan University, Taipei, Taiwan}%

\author{Yang-Ting Chien}
\affiliation{Georgia State University, Atlanta, Georgia, USA}

\author{Gian Michele Innocenti}
\affiliation{CERN, Geneva, Switzerland}%

\author{Marcello Maggi}
\affiliation{INFN Sezione di Bari, Bari, Italy}%

\author{Christopher McGinn}
\affiliation{University of Colorado, Boulder, USA}%

\author{Dennis V. Perepelitsa}
\affiliation{University of Colorado, Boulder, USA}%

\author{Michael Peters}
\affiliation{Massachusetts Institute of Technology, Cambridge, Massachusetts, USA}%

\author{Tzu-An Sheng}
\affiliation{Massachusetts Institute of Technology, Cambridge, Massachusetts, USA}%

\author{Jesse Thaler}
\affiliation{Massachusetts Institute of Technology, Cambridge, Massachusetts, USA}%

\author{Yen-Jie Lee}
\email{yenjie@mit.edu}
\affiliation{Massachusetts Institute of Technology, Cambridge, Massachusetts, USA}%

\date{\today}

\begin{abstract}
The first measurements of energy spectra and substructure of anti-$k_{T}$ jets in hadronic $Z^0$ decays in $e^+e^-$ collisions are presented. The archived $e^+e^-$ annihilation data at a center-of-mass energy of 91.2 GeV were collected with the ALEPH detector at LEP in 1994. In addition to inclusive jet and leading dijet energy spectra, various jet substructure observables are analyzed as a function of jet energy which includes groomed and ungroomed jet mass to jet energy ratios, groomed momentum sharing, and groomed jet radius. The results are compared with perturbative QCD calculations and predictions from the \sherpa, \herwig v7.1.5, \pythia{6}, \pythia{8} and \pyquen event generators. The jet energy spectra agree with perturbative QCD calculations which include the treatment of logarithms of the jet radius and threshold logarithms. None of the event generators give a fully satisfactory description of the data. 
\end{abstract}

%\pacs{}
\maketitle

%\section*{Internal notes}

%To-do's
%\begin{enumerate}
%    \item Add MOD logo on all the plots $\rightarrow$ done!
%    \item \fixme{Go through carefully and add citations}
%    \item what is with the author ordering? (not too important)
%    \item synchronize $Z$ and $Z^0$ (done)
%    \item make png version of the plots (done)
%    \item add event selection cut variation systematics results (Yi) (added some text)
%    \item not all the systematics are done with variation of unfolding matrix, probably want to change the wording. (Yi) (changed wording a bit)
%    \item update acknowledgements if needed (Yen-Jie)
%    \item Fix leading dijet normalization (Yi) (done :O)
%\end{enumerate}

%To-study's
%\begin{enumerate}
%    \item Assign some small value to unfolding to see what will happen (done)
%    \item Hadronic event selection check (done)
%    \item Check other Ben's comments: reweight mean momentum or consituent multiplicity %with fragmentation function and see?
%\end{enumerate}

\clearpage

%\linenumbers

%%%%%%%%%%%%%%%%%%%%%%%%%%%%%%%%%%%%
\section{\label{Section:Introduction}Introduction}
%%%%%%%%%%%%%%%%%%%%%%%%%%%%%%%%%%%%

Jets, sprays of particles originating from the fragmentation and hadronization of a fast-moving quark or gluon, are some of the most useful tools for studying the Quantum Chromodynamics (QCD) and searching for new physics beyond the Standard Model (SM) in high energy colliders. Since the end of the Large Electron Positron Collider (LEP) operation in 2000, significant progress has been made on jet definitions and jet clustering algorithms. Information about the parton shower is not only carried by the four-momentum of jets but also by their internal structure. Therefore, active developments on jet substructure observables have been carried out in order to fully exploit jets~\cite{Abdesselam:2010pt,Salam:2010nqg,Altheimer:2012mn,Larkoski:2017jix}. However, those jet reconstruction and jet substructure algorithms, which are widely used in the data analyses of proton-proton~\cite{Kogler:2018hem} and heavy-ion
collisions~\cite{Connors:2017ptx,Cao:2020wlm,Cunqueiro:2021wls}, have not been used in the most elementary electron-positron $e^+e^-$ collision system. 
Phenomenological models, such as \pythia{}\cite{Sjostrand:2000wi}, \sherpa\cite{Gleisberg:2008ta}, and \herwig\cite{Reichelt:2017hts}, are tuned with hadron level or hadronic event shape observables from LEP experiments and are employed to predict or to describe the jet spectra and jet substructure in more complicated hadron-hadron collision environments. 

%It is beneficial to perform measurements of jet spectra and jet substructure using modern algorithms, compare those results with the theoretical calculations and provide those results as new inputs for the event generator tuning.

Studies of jets, using identical clustering algorithms as those used in high energy hadron colliders such as the Large Hadron Collider (LHC) and Relativistic Heavy-Ion Collider (RHIC), are of great interest.
Unlike hadron-hadron collisions, $e^+e^-$ annihilations do not have beam remnants, gluonic initial state radiation, or the complications of parton distribution functions. The $e^+e^-$ data provide the cleanest test for analytical QCD calculations and phenomenological models that are tuned with hadronic event shapes. Unlike the smooth and steeply falling jet transverse momentum spectra at RHIC and LHC, the inclusive jet and leading jet energy spectra in $e^+e^-$ annihilation are peaked near the half of the collision energy.
Therefore, leading and inclusive jet spectra are sensitive to the jet energy loss from wide-angle radiation that is not clustered into the jet. These spectra provide new tests of the calculation of leading-jet cross-sections, which is sensitive to the treatment of logarithms of the jet radius and threshold logarithms~\cite{Neill:2021std}. Moreover, fully reconstructed jets provide us with an opportunity to inspect quark and gluon fragmentation in great detail on a shower-by-shower basis.
Finally, studies of jet substructure observables and their comparison to modern event generators are of great interest since jet substructure observables are novel tools for jet flavor identification, electroweak boson and top tagging, and studies of the Quark-Gluon Plasma properties at hadron colliders~\cite{Andrews:2018jcm}. Those results can also serve as references for understanding proton-proton collision data and future electron-ion collisions (EIC)~\cite{AbdulKhalek:2021gbh}. 
% add a paragraph about the ee jet spectrum

In this paper, the first measurement of inclusive jet momentum spectra, jet splitting functions, (groomed) jet mass, and groomed jet radius distribution in hadronic $Z^0$ boson decays are presented, with two types of jet selections. The inclusive observables include all jets above an energy threshold and inside a defined acceptance. They are sensitive to higher-order corrections of jet spectra in perturbative QCD. They include low momentum jets in events that are typically associated with soft gluon radiation or random combinatorial jets with hadrons from different partons. On the other hand, the leading dijet observables consider the leading and subleading jet in the event. This type of observable focuses more on the dominant energy flow and is less sensitive to soft radiation.

%%%%%%%%%%%%%%%%%%%%%%%%%%%%%%%%%%%%
\section{\label{Section:ALEPH}ALEPH Detector}
%%%%%%%%%%%%%%%%%%%%%%%%%%%%%%%%%%%%

The ALEPH detector is described in ref.~\cite{Decamp:1990jra}. The central part of the detector is designed for the efficient reconstruction of charged particles. The trajectories of them are measured by a two-layer silicon strip vertex detector, a cylindrical drift chamber, and a large time projection chamber (TPC). Those tracking detectors are inside a 1.5 T axial magnetic field generated by a superconducting solenoidal coil. The charged particle transverse momenta are reconstructed with a resolution of $\delta p_{\rm T}/p_{\rm T} = 6\times 10^{-4} p_{\rm T} \oplus 0.005$ ($p_{\rm T}$ in GeV/c). 

Electrons and photons are identified in the electromagnetic calorimeter (ECAL) situated between the TPC and the superconducting coil. The ECAL is a sampling calorimeter, made by lead plates and proportional wire chambers segmented in $0.9^\circ\times 0.9^\circ$ projective towers. They are read out in three sections in-depth and have a total thickness of around 22 radiation lengths. Isolated photons are reconstructed with a relative energy resolution of $0.18/\sqrt{E}+0.009$ ($E$ in GeV).

The iron return yoke constructed with 23 layers of streamer tubes is also used as the hadron calorimeter (HCAL) for the detection of charged and neutral hadrons. The relative energy resolution for hadrons is $0.85/\sqrt{E}$ ($E$ in GeV). Muons are identified by their pattern in HCAL and by the muon chambers, made by two double-layers of streamer tubes outside the HCAL. 

The information from trackers and calorimeters is combined in an energy flow algorithm~\cite{ALEPH:1994ayc}. This particle flow algorithm outputs a set of charged and neutral particles, called the energy flow objects. They are used in the jet reconstruction in this analysis.

%%%%%%%%%%%%%%%%%%%%%%%%%%%%%%%%%%%%
\section{\label{Section:EventSelection}Dataset and Event Selection}
%%%%%%%%%%%%%%%%%%%%%%%%%%%%%%%%%%%%

The archived $e^+e^-$ annihilation data at a center-of-mass energy of 91.2 GeV were collected with the ALEPH detector at LEP in 1994. To analyze these data, an MIT Open Data format~\cite{Tripathee:2017ybi} was created and validated in  the  two-particle  correlation  function  analysis~\cite{Badea:2019vey}. Hadronic events are selected by requiring the sphericity axis~\cite{Heister:2003aj} to have a polar angle in the laboratory reference frame ($\theta_{\text{lab}}$) between $7\pi/36$ and $29\pi/36$ to ensure that the event is well contained within the detector. At least five tracks having a minimum total energy of 15 GeV are also required to suppress electromagnetic interactions~\cite{Barate:1996fi}. The residual contamination from processes such as $e^+e^-\rightarrow\tau^+\tau^-$ is expected to be less than 0.26\% for these event selections~\cite{Barate:1996fi}. Approximately 1.36 million $e^+e^-$ collisions resulting in the decay of a $Z^0$ boson to quarks are analyzed. 

High-quality tracks from particles are selected using requirements identical to those in previous ALEPH analyses~\cite{Barate:1996fi} and are also required to have a transverse momentum with respect to the beam axis ($p_{\rm T}^{\rm lab}$) above 0.2 GeV/c and $|\cos{\theta_{\text{lab}}}|<0.94$ in the lab frame. Secondary charged particles from neutral particle decays are suppressed by $V^0$ reconstruction in the energy flow algorithm~\cite{Barate:1996fi}. In addition, the total visible energy from energy flow objects is required to be smaller than 200 GeV in order to remove laser calibration events during the run.
  
Event thrust distributions~\cite{Farhi:1977sg} published by the ALEPH Collaboration using a similar dataset~\cite{Heister:2003aj} were successfully reproduced within uncertainties, affirming that the archived data is analyzed properly.  %These factors are parametrized by each track's \pt\ and $\theta$.  They are applied as a weight to each track entering the analysis. \par
  
%%%%%%%%%%%%%%%%%%%%%%%%%%%%%%%%%%%%
\section{\label{Section:MCSamples}Simulated Samples}
%%%%%%%%%%%%%%%%%%%%%%%%%%%%%%%%%%%%
  
Archived \pythia{} 6.1~\cite{Sjostrand:2000wi} Monte Carlo (MC) simulation samples which were produced with the detector conditions during the 1994 run are used to correct for detector effects and to compare with the data. A set of new \pythia{} events are also generated with \pythia{} version 8.303~\cite{Sjostrand:2014zea} at a center-of-mass energy of 91.2~GeV.  The Monash 2013 tune~\cite{Skands:2014pea} is used, with the weak boson exchange and weak single boson processes turned on. In order to select hadronic $Z^0$ decays, pure electroweak events are rejected by requiring at last one hadron in the final state particles. \sherpa samples are produced with version 2.2.5~\cite{Gleisberg:2008ta}, with electron positron events generating 2 to 5 outgoing partons, which are then showered into jets. The coupling constant $\alpha_s (M_Z)$ is set to 0.1188 in the event generation. 
A set of \herwig\cite{Bellm:2015jjp} samples is generated using version 7.2.2. In this sample, 2 to 3 outgoing partons are specified in the hard process, and leptonic decays of $Z^0$ boson are turned off to increase the fraction of hadronic events.  The order of $\alpha$ coupling is set to 2, whereas the colored $\alpha_s$ order is set to 0 in the sample generation. In order to demonstrate the potential modification of the jet spectra and jet substructure observables from jet quenching~\cite{Bjorken:1982tu,ATLAS:2010isq,CMS:2011iwn}, a sample is generated with the \pyquen generator~\cite{Lokhtin:2005px} (version 1.5.3).  The strength of jet  quenching is set to be equivalent to that in a minimum bias sample of PbPb collisions at 5.02 TeV and wide-angle radiation is turned off during the sample generation. %Two subsamples are generated, with and without explicit wide-angle radiations of partons.  The default spectra shown in the results are without wide-angle radiations.

%%%%%%%%%%%%%%%%%%%%%%%%%%%%%%%%%%%%
\section{\label{Section:JetReco}Jet reconstruction and performance}
%%%%%%%%%%%%%%%%%%%%%%%%%%%%%%%%%%%%

%\textbf{[jet reconstruction, calibration and performance]}

% Talk about AK clustering and the e+e- variant
Experimental jets are clustered with the anti-k$_T$ algorithm~\cite{Cacciari:2008gp} with a distance parameter $R = 0.4$, using the \fastjet package~\cite{Cacciari:2011ma} (version 3.3.2). This resolution parameter was chosen since it is widely used in the jet analyses in proton-proton and heavy-ion collisions~\cite{ATLAS:2012tjt,ATLAS:2018gwx,CMS:2021vui,CMS:2016uxf,Dokshitzer:1997in} carried out at the hadron colliders. Moreover, the chosen value also gives us an opportunity to examine the shower from quarks in detail. The distance measure $d_{ij}$ between jets $i$ and $j$ and $d_{iB}$ between jet $i$ and the beam is defined using the opening angle $\theta_{ij}$:
\begin{align}
    d_{ij} &= \min(E_i^{-2}, E_j^{-2}) \dfrac{1 - \cos\theta_{ij}}{1 - \cos R}\nonumber\\
    d_{iB} &= E_i^{-2},\nonumber
\end{align}
where $E_i$ is the energy of the $i$-th jet and $\theta_{ij}$ is the opening angle of the $i$-th and $j$-th jets.  The termination of the clustering process is defined by the distance between the jet and the beam direction, $d_{iB}$.
In the data analysis, energy flow objects which are reconstructed using the tracker and calorimeter information, are used for jet reconstruction.
The mass of the objects are assumed to be pion for charged hadron candidates, and massless for photon candidates.
%\fixme{we talked about energy flow in section 3 already, though perhaps it's fine to repeat information for clarity} 
Generator-level jets are clustered by considering all visible final state particles by the ALEPH detector (i.e., excluding neutrinos).  Reconstructed-level jets are clustered with all energy flow candidates reconstructed by the tracker and calorimeters. In order to avoid overlapping with the beam pipe, only jets between $0.2\pi < \theta_\text{jet} < 0.8\pi$ are considered in this measurement, where $\theta_\text{jet}$ is the angle between jet momentum and beam axis.

% Calibration strategy: multi-stage
The jets are calibrated following a multi-stage strategy.  In the first stage, simulated jet responses are corrected by using generated jets, clustered with visible final state particles with a distance parameter of $R = 0.4$, as the reference.  The correction factor is derived as a function of jet 3-momentum $p_\text{jet}$ in 18 bins of jet direction $\theta_\text{jet}$.  The closure in the simulated sample is seen to be better than 0.5\%.%refer to the arxiv post

The jets in data are corrected with the first stage correction from simulation before deriving residual corrections characterizing the difference between data and simulation.  In the second stage, there are two steps in the residual correction: relative residual correction, which equalizes the jet response between the two halves of the detector, and absolute residual correction, which calibrates the jet to an absolute scale. The jet relative correction is performed using the leading jet distribution in different bins of $\theta_\text{jet}$ and comparing the two sides of the detector.  This procedure aims to unify the jet energy scale in data between the two sides of the detector.  The correction is typically of scale up to 0.5\% in data, while the result on the simulated sample with the same procedure is consistent with 0.

A cross-check is carried out using the momentum balance of dijets in bins of third jet activity related to the leading dijet: $E_3/(E_1+E_2)$.  The residual scale is derived by extrapolating the third jet activity to zero.  Due to the larger uncertainty induced by the extrapolation procedure, this procedure is not adopted as nominal.   The corrections between the cross-check and the nominal one are consistent within the uncertainty.

The absolute correction of jets is performed using multi-jet events with an invariant mass around the $Z^0$ peak.  Both the simulation correction and the residual relative correction are applied before the derivation of the absolute correction.  The overall scale is fitted using up to $N$ high energy jets (with $N$ ranging from 3 to 9), and with different requirements on the energy of the $(N+1)$-th leading jet from 1 to 5 GeV as a handle to control potential bias from the method.  If there are less than $N$ jets reconstructed in the event, the $(N+1)$-th jet has 0 energy and the event is included.  The correction is derived in bins of $\theta_\text{jet}$ and is up to 0.3\%.

% Jet energy resolution difference between data and MC
The jet energy resolution in simulation is measured using an empirical function, in bins of $\theta_\text{jet}$:
\begin{align}
    \sigma_\text{jet} = \sqrt{a_0 + \dfrac{a_1}{p_\text{jet}} + \dfrac{a_2}{p_\text{jet}^2} + \dfrac{a_3}{p_\text{jet}^3}},\nonumber
\end{align}
where $a_0$ to $a_3$ are parameters to be fitted.  Depending on the direction of the jet, $a_0$ ranges from 0.005--0.010 GeV$^2$, $a_1$ from 0.58--0.72 GeV$^3$, $a_2$ up to 0.62 GeV$^4$, while $a_3$ is negligible.

The difference between jet energy resolution between data and simulation is measured using the momentum balance of the leading dijet in an event.  The magnitude of the third-leading jet is used as a handle to understand potential systematic bias from the method, and it is constrained to be less than 3 GeV in the final result.  The difference is characterized as a scale factor, in bins of $\theta_\text{jet}$.  The resolution in data is found to be up to 5\% worse than that in simulation, depending on the direction of the jet~\cite{Chen:2021iyj}.

%%%%%%%%%%%%%%%%%%%%%%%%%%%%%%%%%%%%
\section{\label{Section:Analysis}Analysis}
%%%%%%%%%%%%%%%%%%%%%%%%%%%%%%%%%%%%

\subsection{Jet Grooming}

% The soft drop algorithm and the modification to adapt for e+e-
The soft drop algorithm~\cite{Larkoski:2014wba,Dasgupta:2013ihk} is proposed as a way to define a set of theoretically well-controlled observables that are sensitive to structures inside a jet and has been used in various QCD measurements in proton-proton and heavy-ion collisions~\cite{CMS:2017qlm,ALICE:2019ykw, STAR:2020ejj,ATLAS:2019mgf,CMS:2018ypj,CMS:2018fof}.  It is characterized by two free parameters $z_\text{cut}$ and $\beta$.  The constituents of a jet are first reclustered using the Cambridge-Aachen algorithm~\cite{Dokshitzer:1997in,Wobisch:1998wt}, and then the clustering history is traced, and the algorithm terminates when the soft drop condition is satisfied:
\begin{align}
    z \equiv \dfrac{\min(E_1, E_2)}{E_1 + E_2} \geq z_\text{cut} \left(\dfrac{\theta_{12}}{R}\right)^\beta,\nonumber
\end{align}
where $E_1$ and $E_2$ are the energies of the two branches, and $\theta_{12}$ is their opening angle.  Note the definition is adapted to using the energy and opening angle (a true Great-Circle Distance), as opposed to using the transverse energy and $\Delta R \equiv \sqrt{(\Delta\eta)^2+(\Delta\phi)^2}$ as its counterpart in a hadronic collider environment.  When the algorithm terminates, the $z$ is defined as \zg~\cite{Larkoski:2017bvj}, and the opening angle is defined as \Rg.

In this measurement, in addition to the jet energy spectrum, the \zg and \Rg spectra in bins of jet energy for $z_\text{cut} = 0.1, \beta = 0$ are reported.  For the \zg and \Rg distributions, the distributions are self-normalized in order to decouple effects coming from jet energy migration and the rest.

\subsection{Unfolding}

% The Bayes iterative unfolding method
In order to mitigate smearing effects from the detector, an unfolding is performed on the measured spectra.  The unfolding is done with a Bayes' iterative method based on ref.~\cite{Dagostini:1994fjx}, as implemented in the \textsc{RooUnfold} package.  The method applies an iterative method with a corrected error calculation which decreases sensitivity to the prior knowledge used in the Bayes' formulation.

% Iteration optimization
The ideal number of iterations for the unfolding is obtained with a two-step procedure.  The generator level distribution from the simulation is first combined with the smearing matrix to produce the idealized detector level distribution.  Pseudodatasets are then generated using this distribution with the number of statistics compatible with that of the data.  The ideal number of iterations can then be obtained by comparing the unfolded distribution with the generator level distribution.  In the second step, data is first unfolded using the number of iterations found from the first step, and the whole procedure is repeated by using the unfolded distribution in place of the generator level distribution.  The numbers range from 7 to 17 and the exact numbers can be found in ref.~\cite{Chen:2021iyj}.

\begin{figure}
    \centering
    \includegraphics[width=0.45\textwidth]{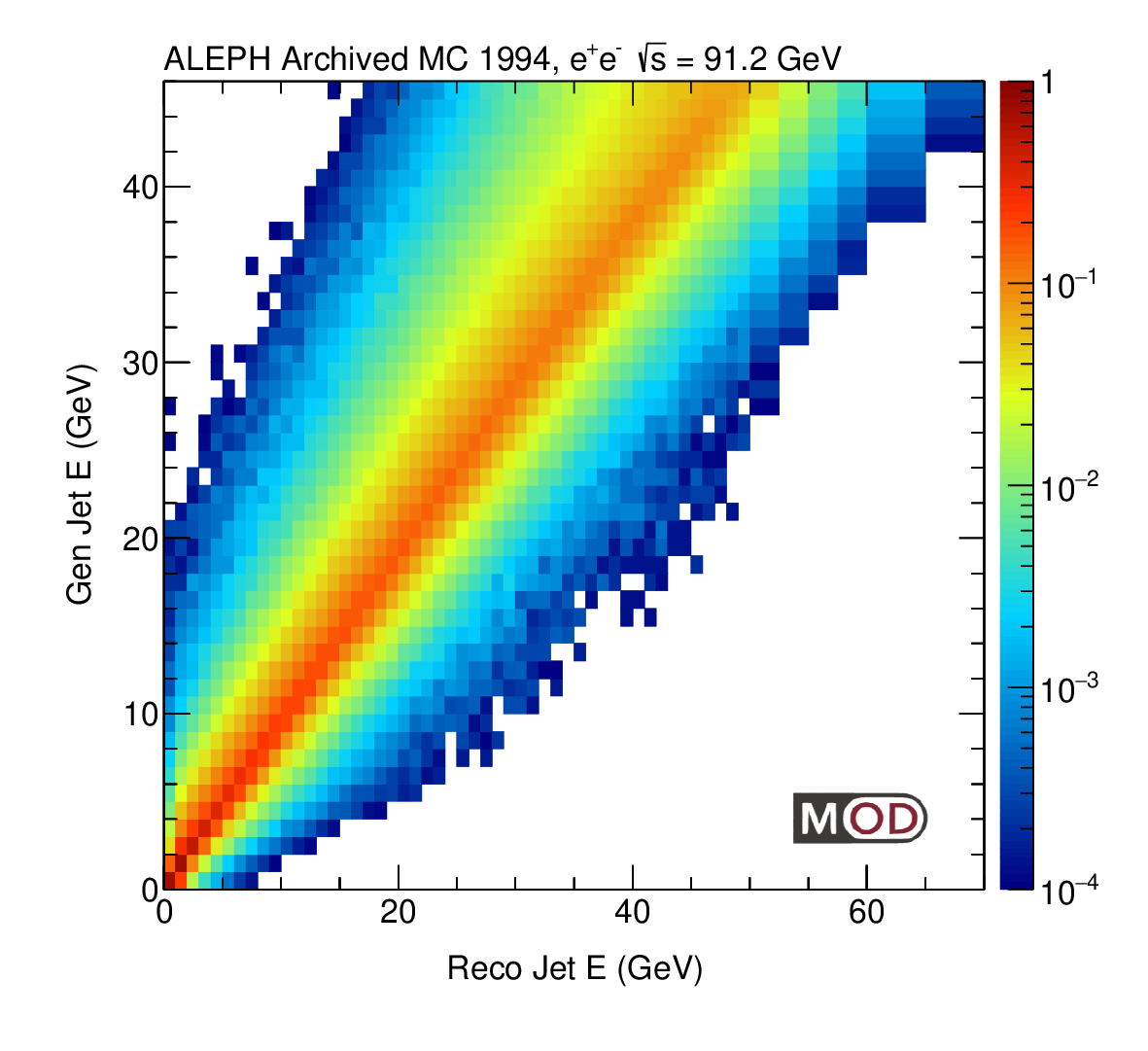}
    \caption{Response matrix for jet $E$.  The matrix is row-normalized: the color scale indicates the probability density of a given generator level jet being smeared into a different reconstructed level jet.}
    \label{Figure:ResponseJetP}
\end{figure}

% Description of jet energy unfolding
The response matrix for the jet energy measurement is shown in Figure~\ref{Figure:ResponseJetP}.  The measurements are done in bins of 1~GeV up to 45 GeV, after which the bin sizes are progressively larger in order to accumulate enough entries in the tail. In Figure~\ref{Figure:ResponseJetP}, the matrix is row-normalized: the color scale indicates the probability density of a generated (``Gen'') jet of given energy being smeared into different reconstructed (``Reco'') jet energy.  A tight correlation between the reconstructed jet and the matched truth-level jet is observed. No significant contribution from outliers is present in the correlation matrix.  The final result for this measurement starts from 10 GeV based on studies of jet energy scales.  The bins below 10 GeV are included in the unfolding procedure as underflow bins, but they are not part of the final result.
%\begin{figure}
%    \centering
%    \includegraphics[width=0.45\textwidth]{letter/plots/response/MatrixLeadingDiJet.png}
%    \caption{Response matrix for leading dijet jet $E$.  The matrix is row-normalized: the color scale indicate the probability density of a given generator level (``Gen'') jet being smeared into different reconstructed level (``Reco'') jet. \textbf{REMOVE}}
%    \label{Figure:ResponseLeadingDiJet}
%\end{figure}

\begin{figure*}
    \centering
    \includegraphics[width=0.9\textwidth]{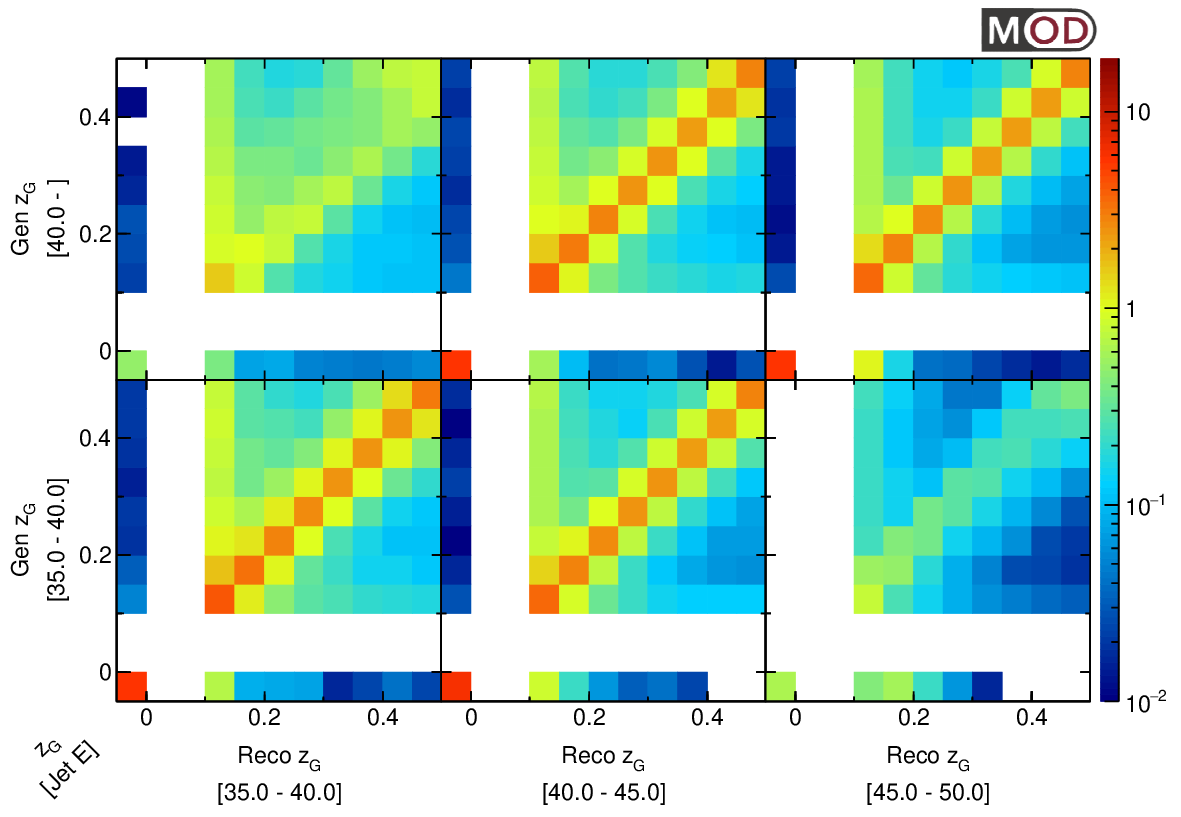}
    \caption{Partial response matrix for $z_G$, in bins of jet $E$.  Each block is the sub-matrix for the indicated jet $E$ range.  The matrix is row-normalized across all jet $E$ ranges: the color scale indicates the probability density of a given generator level (``Gen'') jet being smeared into a different reconstructed level (``Reco'') jet.}
    \label{Figure:ResponseJetZG}
\end{figure*}

\begin{figure*}
    \centering
    \includegraphics[width=0.9\textwidth]{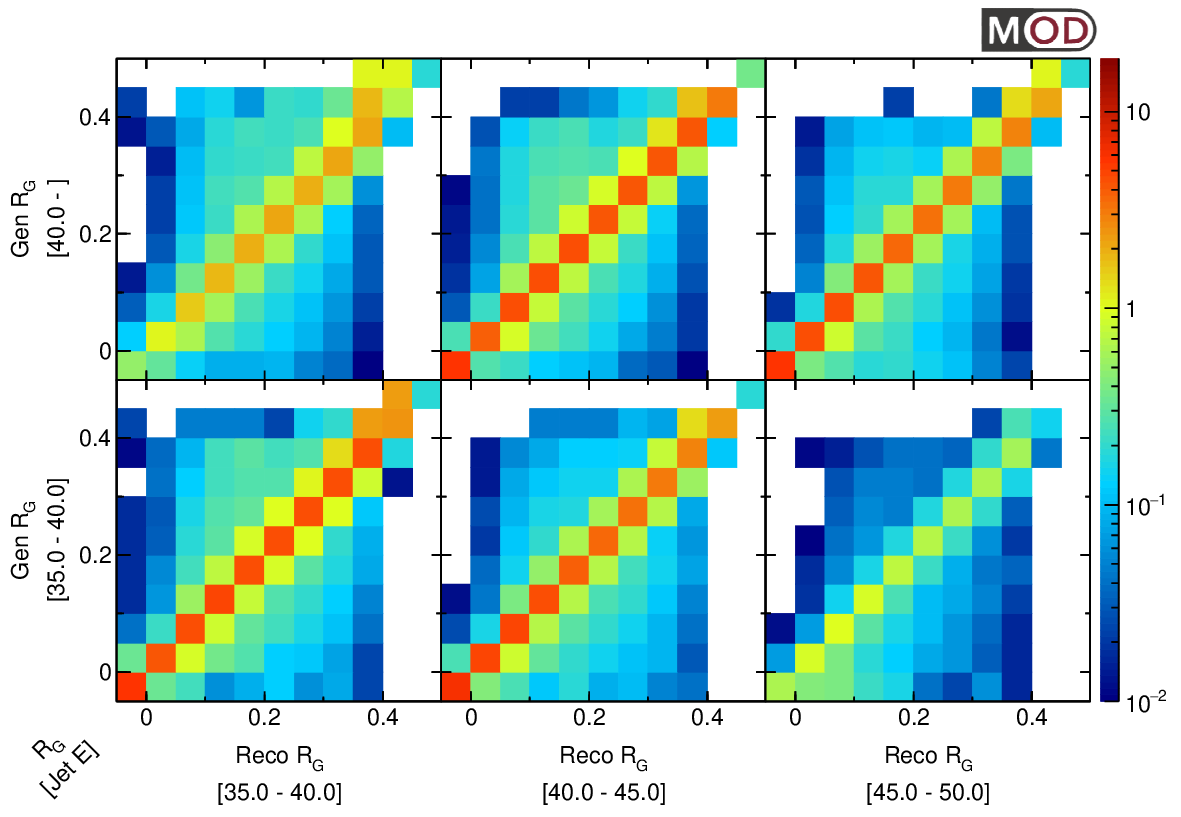}
    \caption{Partial response matrix for $R_G$, in bins of jet $E$.  Each block is the sub-matrix for the indicated jet $E$ range.  The matrix is row-normalized across all jet $E$ ranges: the color scale indicates the probability density of a given generator level (``Gen'') jet being smeared into a different reconstructed level (``Reco'') jet.}
    \label{Figure:ResponseJetRG}
\end{figure*}

\begin{figure*}
    \centering
    \includegraphics[width=0.9\textwidth]{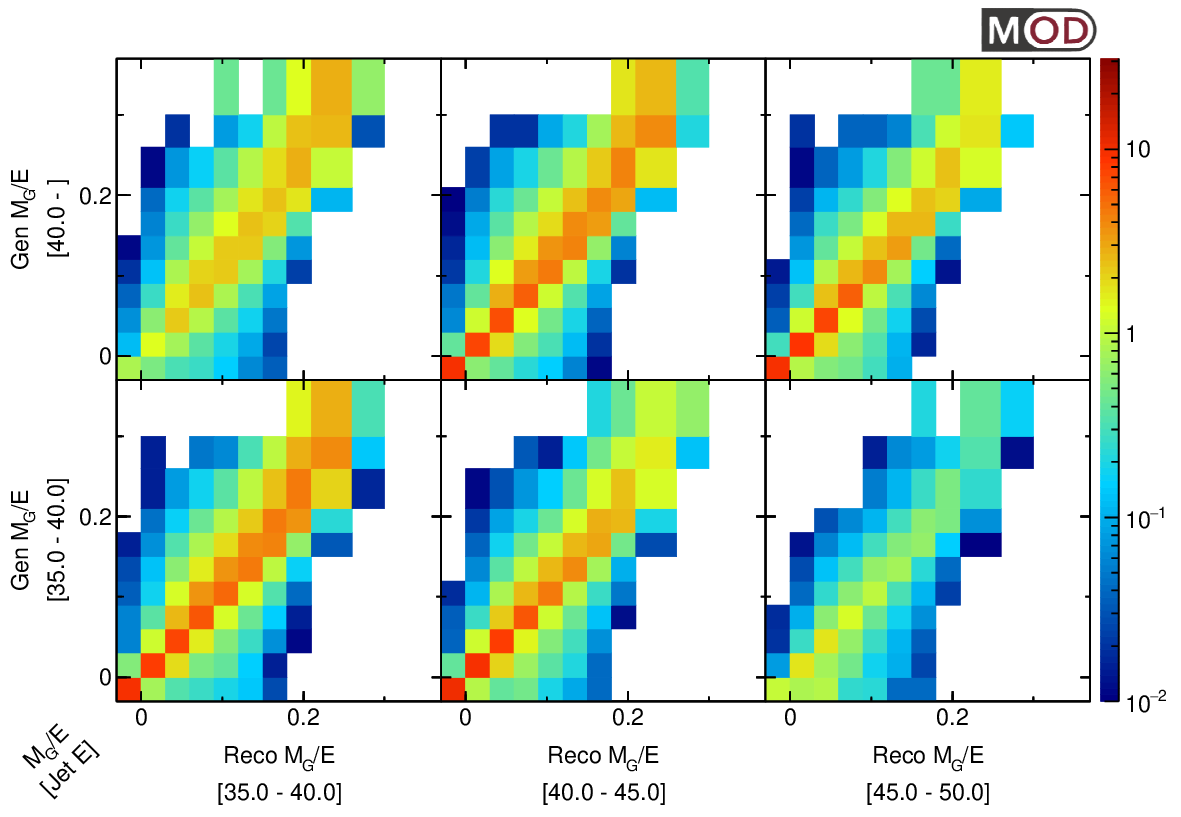}
    \caption{Partial response matrix for $M_G/E$, in bins of $E$.  Each block is the sub-matrix for the indicated $E$ range.  The matrix is row-normalized across all $E$ ranges: the color scale indicates the probability density of a given generator level (``Gen'') jet being smeared into a different reconstructed level (``Reco'') jet.}
    \label{Figure:ResponseJetMGE}
\end{figure*}

\begin{figure*}
    \centering
    \includegraphics[width=0.9\textwidth]{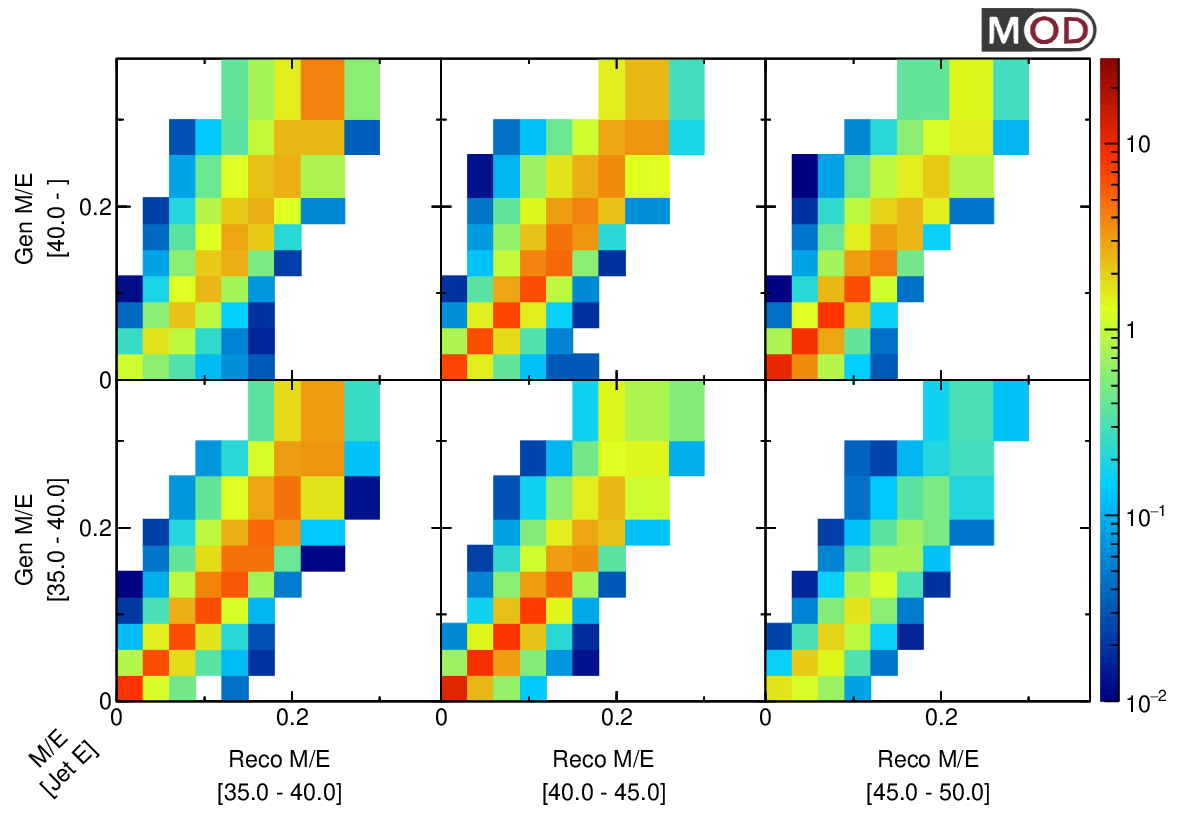}
    \caption{Partial response matrix for $M/E$, in bins of $E$.  Each block is the sub-matrix for the indicated $E$ range.  The matrix is row-normalized across all $E$ ranges: the color scale indicates the probability density of a given generator level (``Gen'') jet being smeared into a different reconstructed level (``Reco'') jet.}
    \label{Figure:ResponseJetME}
\end{figure*}

% The need for 2D unfolding for groomed observables due to jet energy bin migration
Due to non-negligible migration of jet energy, the substructure observables are unfolded together with the jet energy $E$: $(E, z_G)$, $(E, R_G)$, $(E, M/E)$ and $(E, M_G/E)$.  The jet energy is binned in 5~GeV bins until the kinematic limit, after which the bin size is increased due to the available number of entries.  Part of the response matrices are shown in Figs.~\ref{Figure:ResponseJetZG}, \ref{Figure:ResponseJetRG}, \ref{Figure:ResponseJetMGE}  and \ref{Figure:ResponseJetME}.  The full matrices are too large to be properly included in the main paper body.  The matrices are row-normalized.  For most of the matrix, within each block of jet energy, the response is mostly diagonal, and there are significant contributions in the off-diagonal blocks due to jet energy migration.  For \zg and \Rg, there is an off-diagonal contribution within each block at a few percent level.  These correspond to cases where the detector smearing causes the soft drop algorithm to terminate at a different point in the declustering history.  At very low energy below 10 GeV (not shown in the partial matrices), the response is observed to be less diagonal.  This indicates that there might be a significant contribution of jets of combinatorial origin.

%\fixme{What do we do when there is no simulation in some of the bins during unfolding? (Check what will happen if we assign some small value)}

% Statistical uncertainty cross check to validate unfolding uncertainty
The statistical uncertainty from the unfolding machinery is validated using pseudoexperiments.  A set of 250 pseudodatasets are generated by combining the unfolded data distribution and the smearing matrix.  Each of the pseudodatasets is unfolded, and the pull distribution between the unfolded pseudodataset and the input distribution is examined.  For all cases, the width of the pull distribution is consistent with 1.

\subsection{Leading dijet selection}

There is no detector acceptance close to the beamline, and therefore in order to measure only the event-wide leading and subleading jet, referred to as ``leading dijet'' in the later text, an additional selection is designed.  The total energy visible either inside the acceptance $0.2\pi < \theta < 0.8\pi$ or within radius $R$ of any jet above 5 GeV is calculated, and events with total visible energy above 83 GeV are selected to ensure a purity of 99\% of the events having leading dijet inside the acceptance.  The efficiency of such a selection is about 60\%.  A correction is derived using simulated events to correct for effects on the spectra due to this extra selection.

%%%%%%%%%%%%%%%%%%%%%%%%%%%%%%%%%%%%
\section{\label{Section:Systematics}Systematic uncertainties}
%%%%%%%%%%%%%%%%%%%%%%%%%%%%%%%%%%%%

The majority of systematic uncertainties are evaluated by unfolding the data with alternative versions of smearing matrices that encompass the systematic variations we would like to examine, with a few exceptions, as discussed individually below.  Different sources are added in quadrature.  These sources are discussed in the following paragraphs:

\textbf{Systematics related to hadronic event selections}

The requirement on the number of charged particles passing all track selections is changed from 5 to 6 in order to check the systematics related to the hadronic event selection.  The full analysis chain is rerun with the new selection, and the results are compared to the nominal selection.  The deviation is negligible ($< 0.1\%$).

\textbf{Systematics related to jet reconstruction}

% Jet energy scale uncertainty
The residual jet energy scale (``JEC'') is varied by $\pm 0.5\%$ which is the maximum magnitude of the residual corrections.  This uncertainty source ranges from a few percent up to 20\% in jet energy spectra, and it is the dominating source of uncertainty at large $E_\text{jet}$.  For jet substructure results, due to the self-normalization, jet energy scale uncertainty, which manifests itself mostly as an overall normalization in each energy bin, is small.

% Jet energy resolution uncertainty
The jet energy is smeared by $\pm2.5\%$, the difference between data and MC simulation extracted from dijet balance studies, to construct alternative response matrices in order to evaluate the uncertainty associated with the resolution (``JER'') difference between data and simulation.  It is the dominating source in the mid-energy region and ranges from a few percent up to 15\%.

% MC reweighting uncertainty
%In the analysis, the archived simulation is used as the nominal.  The uncertainty coming from this choice of simulation is estimated.  The simulated sample is reweighted before building the smearing matrix and repeating the full analysis.  Several different types of reweighting are considered: those based on the particle multiplicity with different energy threshold (0.5 GeV, 1.0 GeV and 1.5 GeV), and those based on the substructure observable $z_G$ and $R_G$.  For each of them, ratio to other generator is used as a guide to how much to vary.  The unfolded results are seen to differ by up to a few percent for the observables, and the difference is quoted as the systematic uncertainty.
The corrections to detector effects are extracted from the archived Monte Carlo.  We study the uncertainty coming from this choice of simulation with the following procedure.  We first reweight the simulated sample to match predictions on certain observables from other event generators, such as \herwig and \sherpa, before building the smearing matrix and repeating the whole analysis.  We considered the following target observables in the reweighting studies: (1) reweight the jet constituent multiplicity with different energy thresholds on the constituents (0.5 GeV, 1.0 GeV, and 1.5 GeV); (2) reweight MC to match the substructure observable $z_G$ and $R_G$.  The unfolded results with reweighted MC changed by up to a few percent, and the maximum difference from the nominal result is quoted as the systematic uncertainty.

% Substructure extra uncertainty
%There are additional uncertainties related to the modeling of substructure observables.  Two different types are considered: those coming from the imperfect modeling of angular resolution of subjets and those from the energy resolution.  For the angular resolution, subjets are smeared randomly based on the observed simulated resolution.  The choice is conservative but the overall effect is small.  The energy of the subjets are also smeared based on twice the energy resolution scale factor obtained for the jets.  The total uncertainty from these two sources amounts to 3\% for $z_G$ and $R_G$, and 5\% for jet mass.  The uncertainty from substructure modeling and MC reweighting are grouped under the ``Model'' category in the summary plots.

For the angular and energy resolution of the subjets, the associated uncertainties are determined in the following way.
The directions of the subjet momenta are smeared randomly based on the observed simulated resolution.  Although this choice is conservative,  the overall effect is small.  The energy of the subjets is also smeared based on twice the nominal energy resolution scale factor (+5\%) obtained for the jets.  The total uncertainty from these two sources amounts to 3\% for $z_G$ and $R_G$, and 5\% for jet mass.  The total uncertainties from substructure modeling and MC reweighting are shown in the ``Model'' category in the summary plots.

% Fake jets
Contributions from fake or combinatorial jets are estimated by matching reconstructed jets in simulation to generator-level jets.  The fraction of unmatched jets indicates the amount of potential fake jets.  It ranges from 0--5\% depending on the jet energy for inclusive jets.

% Other checks
A check on the results has been carried out by comparing the measured spectrum from each side of the detector ($0.2\pi < \theta_\text{jet} < 0.5\pi$ and $0.5\pi < \theta_\text{jet} < 0.8\pi$).  The spectra are consistent with each other.

\textbf{Systematics related to unfolding}

% Choice of prior + iterations
We examine the sensitivity to the choice of prior which is part of the input to the unfolding.  The difference between a flat prior and a prior with the generator level distribution from the simulation is taken as the uncertainty.  It is grouped (``Unfold'') with the uncertainty associated with the number of iterations done in the unfolding process.  The number of iterations is varied from the determined ideal number of iterations to that optimized on a separate sample (ie., through Monte Carlo studies), and the difference in the unfolded distribution is taken as uncertainty.

% Unfolding method: SVD
In order to assess for potential bias in the unfolding method itself, an alternative unfolding method, the singular value decomposition (SVD) method~\cite{Hocker:1995kb}, is used.  The SVD unfolding method employs a regularization parameter, and the ideal parameter is determined using a method similar to that used to determine the ideal number of iterations in the nominal unfolding method.  The difference between the SVD and the nominal is quoted as a systematic uncertainty.

% Unfolding nonclosure in MC
A final source of systematic uncertainty related to the unfolding is the observed nonclosure on simulated events.  This source of uncertainty is negligible for jet $E$ and sizable, but still subdominant, for \zg and \Rg when $E$ is small.

\textbf{Systematics related to leading dijet selection}

The uncertainty on the extra leading dijet selection is evaluated by changing the thresholds on the total visible energy to those corresponding to a leading dijet purity of 98\% and 99.5\%.  The difference to nominal is taken as the systematic uncertainty.  The effect from potential imperfect modeling of the resolution of the total visible energy is investigated by smearing the quantity event by event and repeating the analysis.  The difference is added in quadrature to the scale uncertainty.

Uncertainty that arises from the correction factor is evaluated through a reweighting procedure on the simulated sample.  The events are weighted by the ratio of jet spectra for each jet between the nominal unfolded data and simulation, and the weighted simulated sample is used to derive the alternative correction factor.  The difference is quoted as uncertainty.

\begin{figure}
    \centering
    \includegraphics[width=0.45\textwidth]{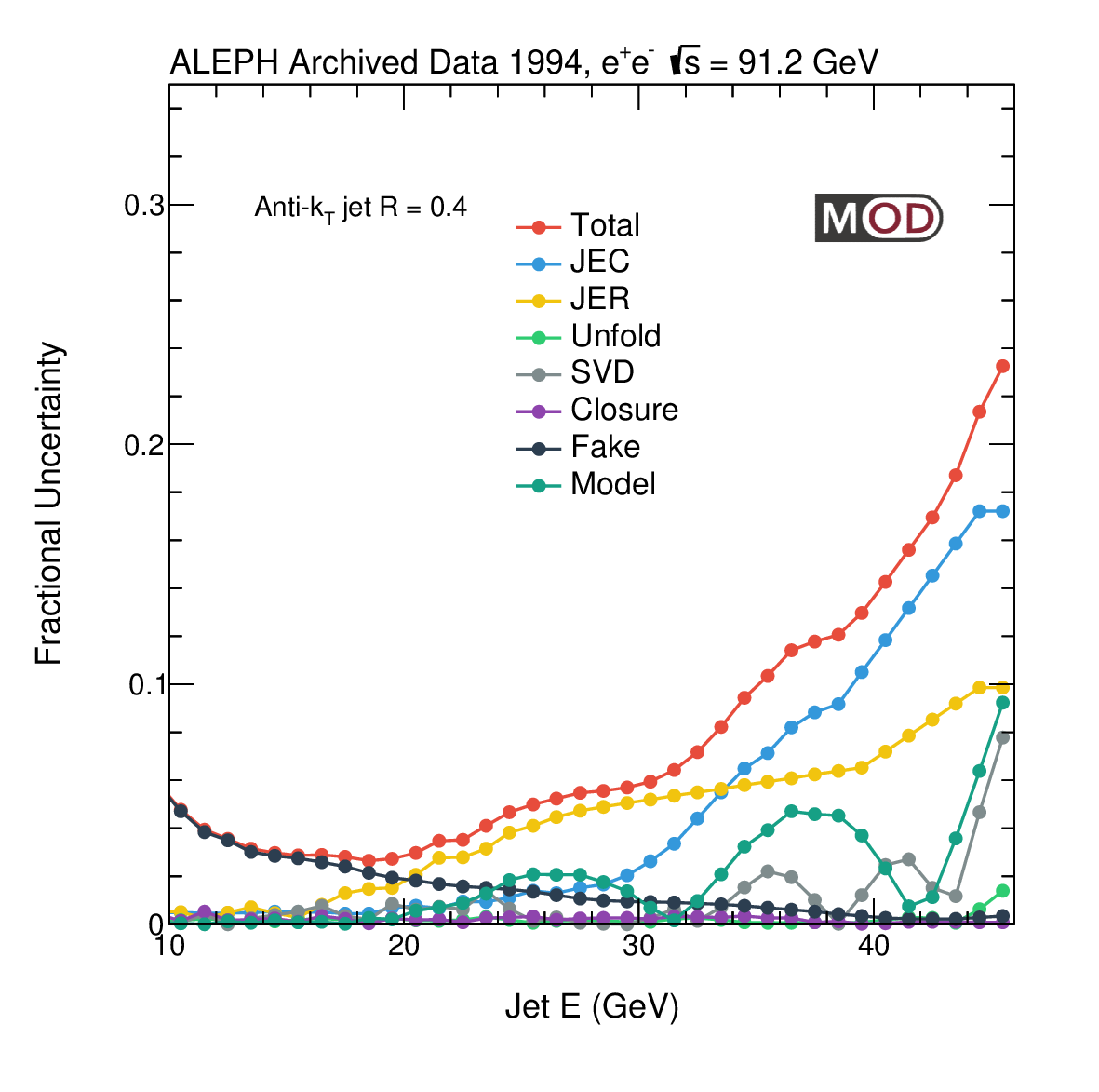}
    \caption{Summary of the size of relative systematic uncertainty from different sources for jet $E$.}
    \label{Figure:SystematicsJetE}
\end{figure}

\begin{figure}
    \centering
    \includegraphics[width=0.45\textwidth]{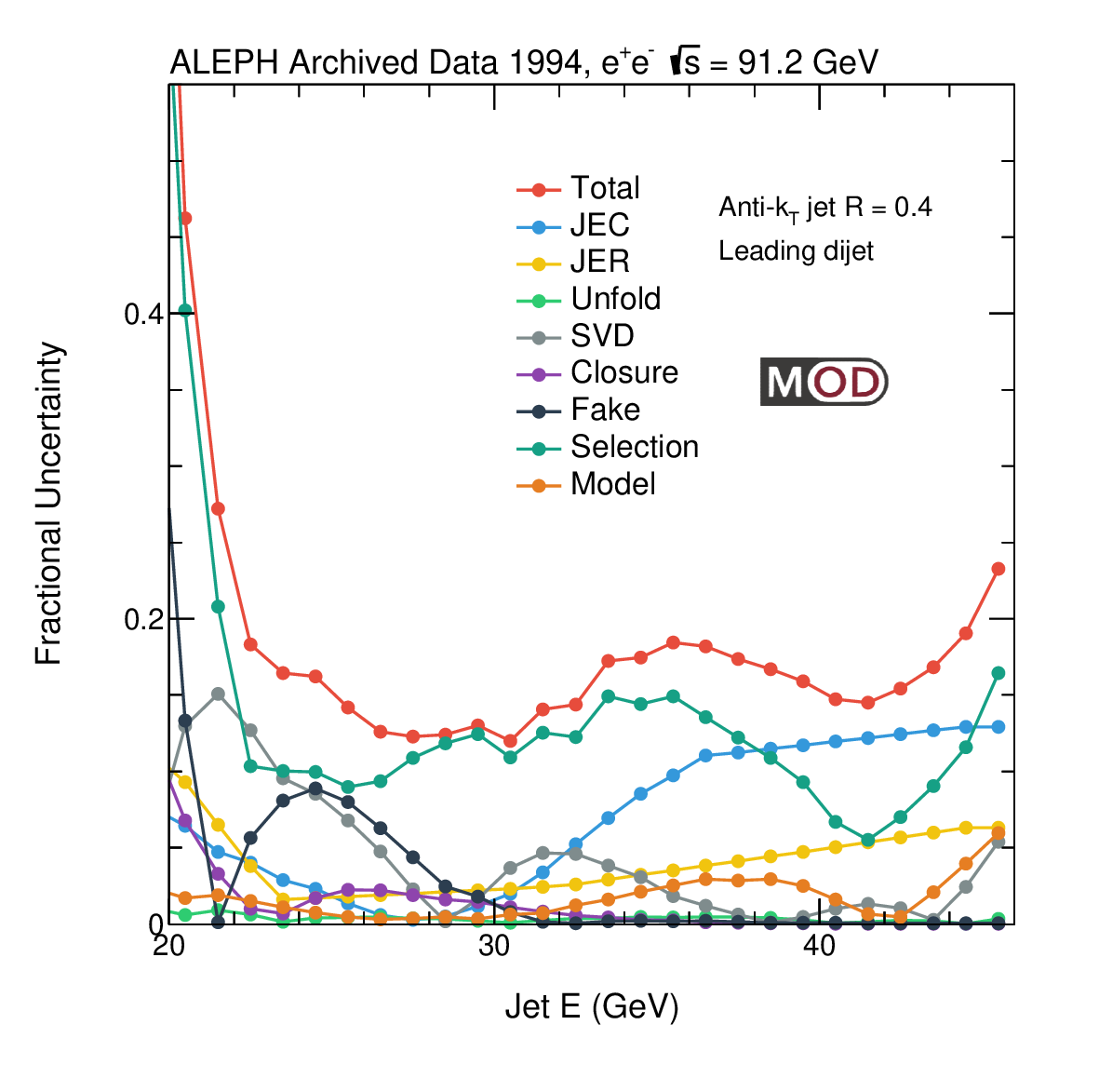}
    \includegraphics[width=0.45\textwidth]{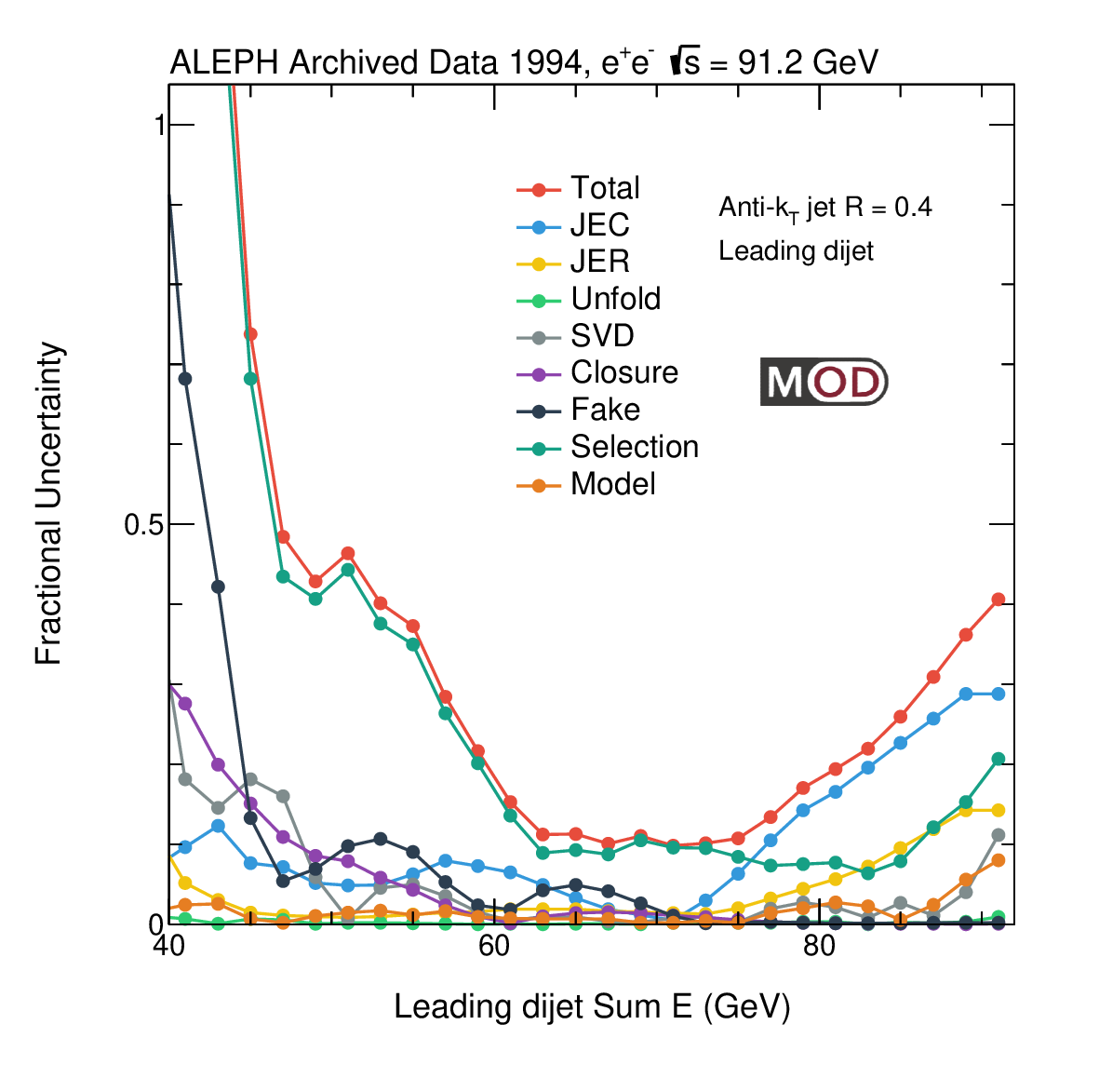}
    \caption{Summary of the size of relative systematic uncertainty from different sources for each of the two jet $E$ in leading dijet (left) and leading dijet total energy (right).}
    \label{Figure:SystematicsLeadingDiJetE}
\end{figure}

\begin{figure*}
    \centering
    \includegraphics[width=0.9\textwidth]{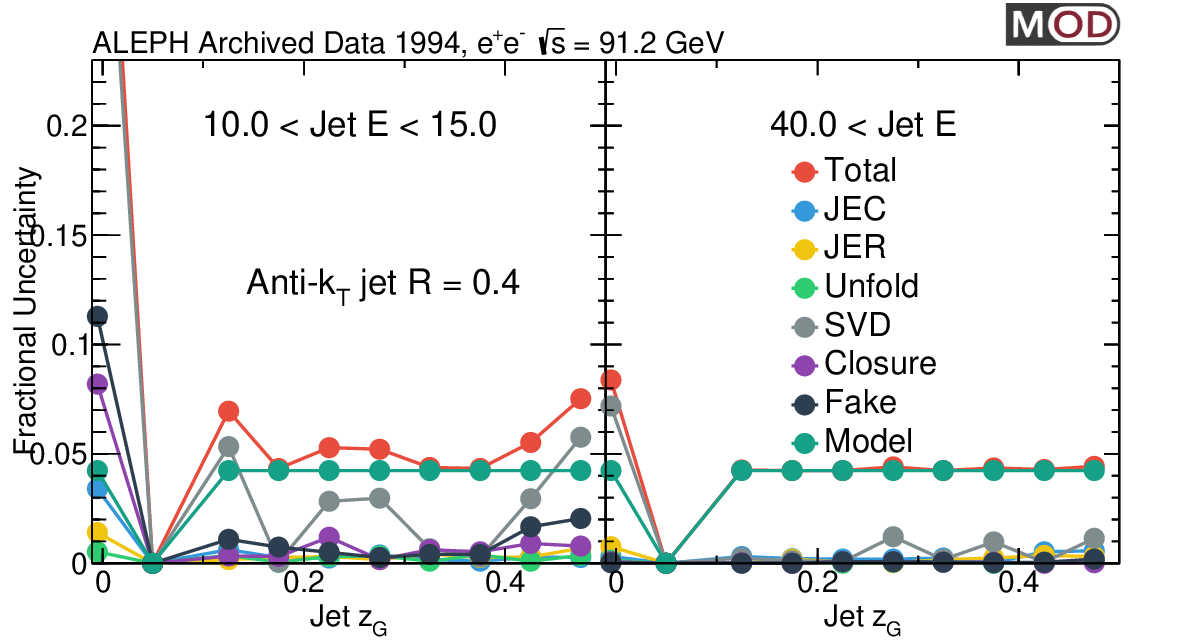}
    \caption{Summary of the size of relative systematic uncertainty from different sources for $z_G$, in bins of jet $E$ from 10--15~GeV (left panel) to above 40~GeV (right panel).}
    \label{Figure:SystematicsJetZG}
\end{figure*}

\begin{figure*}
    \centering
    \includegraphics[width=0.9\textwidth]{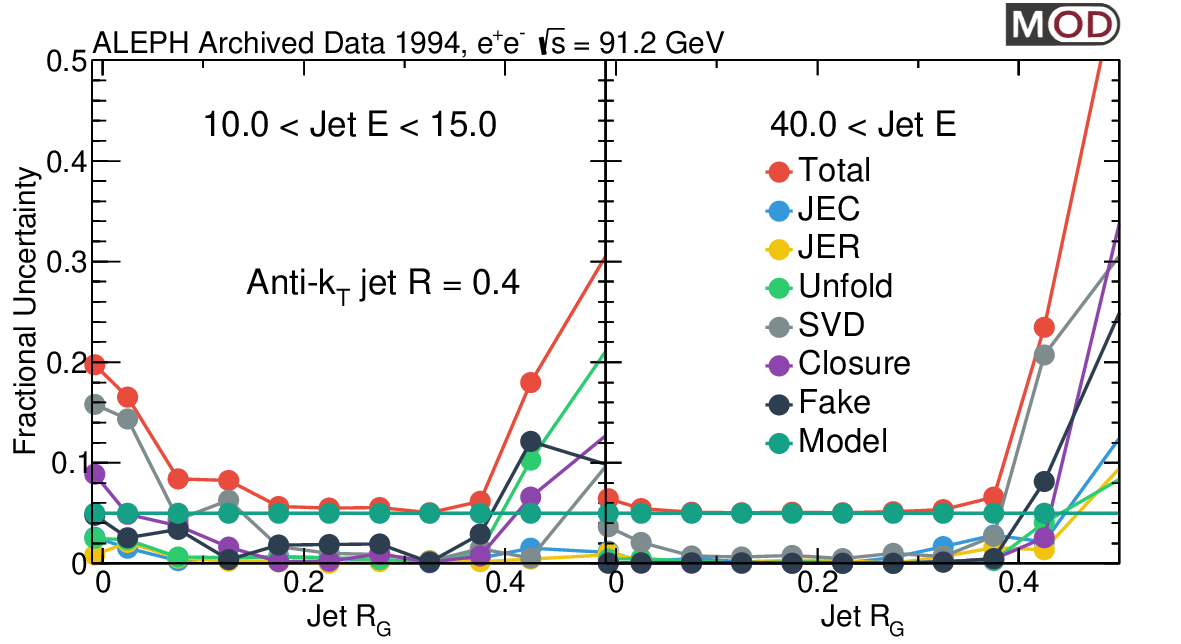}
    \caption{Summary of the size of relative systematic uncertainty from different sources for $R_G$, in bins of jet $E$ from 10--15~GeV (left panel) to above 40~GeV (right panel).}
    \label{Figure:SystematicsJetRG}
\end{figure*}

\begin{figure*}
    \centering
    \includegraphics[width=0.9\textwidth]{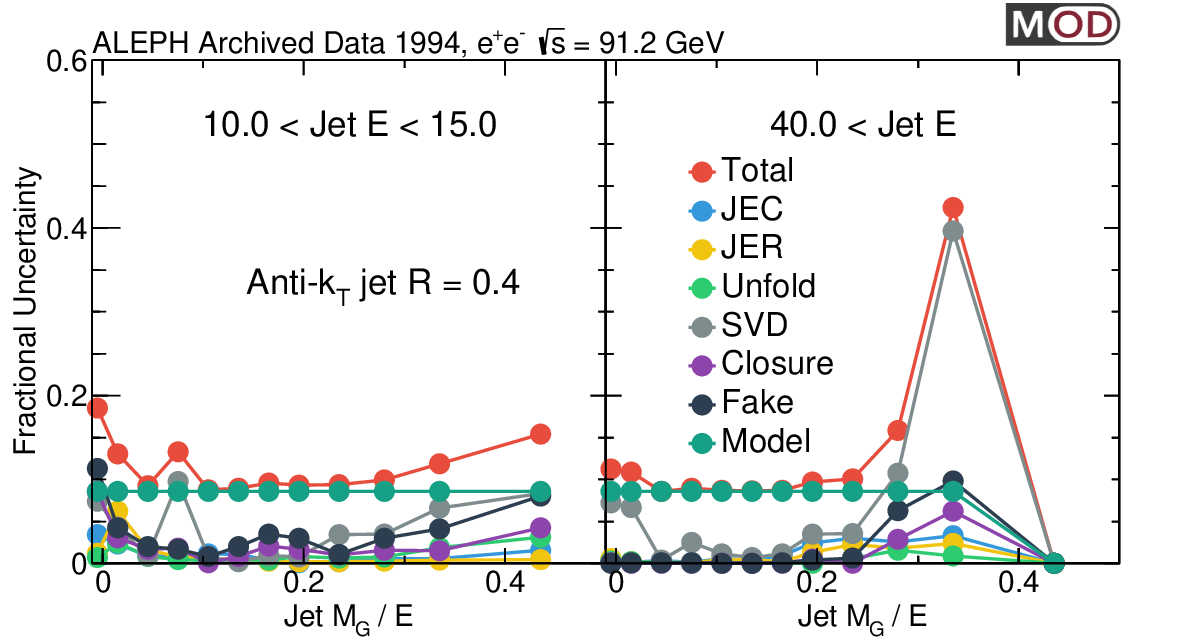}
    \caption{Summary of the size of relative systematic uncertainty from different sources for $M_G/E$, in bins of jet $E$ from 10--15~GeV (left panel) to above 40~GeV (right panel).}
    \label{Figure:SystematicsJetMGE}
\end{figure*}

\begin{figure*}
    \centering
    \includegraphics[width=0.9\textwidth]{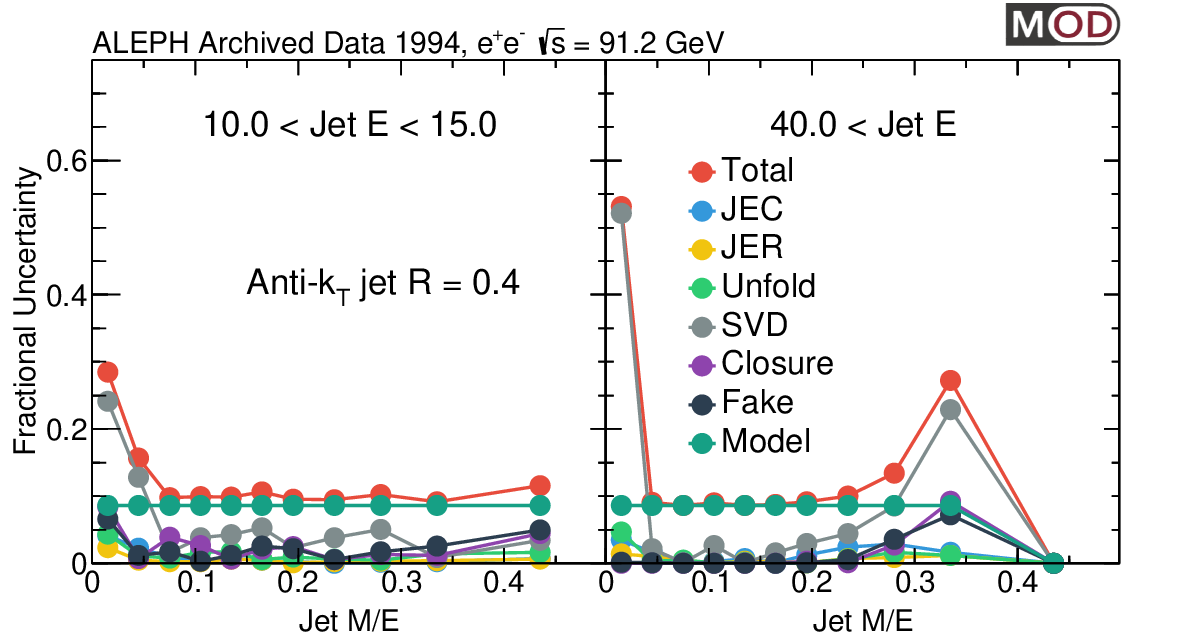}
    \caption{Summary of the size of relative systematic uncertainty from different sources for $M/E$, in bins of jet $E$ from 10--15~GeV (left panel) to above 40~GeV (right panel).}
    \label{Figure:SystematicsJetME}
\end{figure*}

A summary of the systematic uncertainties are shown in Fig.~\ref{Figure:SystematicsJetE} for jet $E$, Figs.~\ref{Figure:SystematicsLeadingDiJetE} for leading dijet, and Figs.~\ref{Figure:SystematicsJetZG}, \ref{Figure:SystematicsJetRG}, \ref{Figure:SystematicsJetME}, and \ref{Figure:SystematicsJetMGE} for the jet substructure observables.

\clearpage

%%%%%%%%%%%%%%%%%%%%%%%%%%%%%%%%%%%%
\section{\label{Section:Results}Results and discussions}
%%%%%%%%%%%%%%%%%%%%%%%%%%%%%%%%%%%%

The fully corrected jet spectra and jet substructure observables are presented and compared with predictions from \pythia{6} (red), \pythia{8} (blue), \herwig (green), and \sherpa (purple) event generators. The results are also compared to analytical calculations with perturbative QCD. Finally, predictions from the PYQUEN (gray) event generator, which added jet quenching effect to the simulated $e^+e^-$ events, are also overlapped to illustrate the possible modifications due to the presence of a strongly interacting medium.
\begin{figure}
    \centering
    \includegraphics[width=0.45\textwidth]{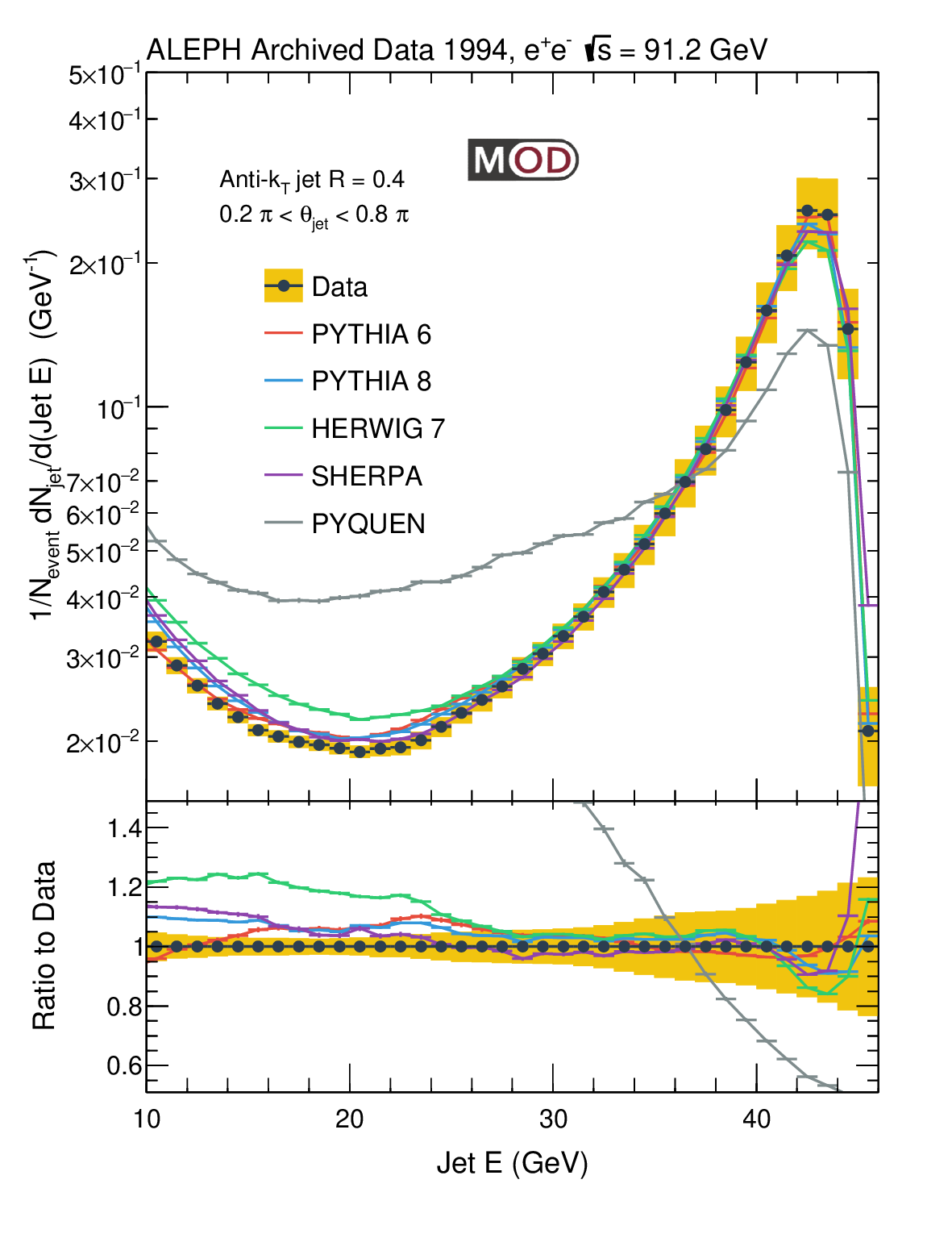}
    \includegraphics[width=0.45\textwidth]{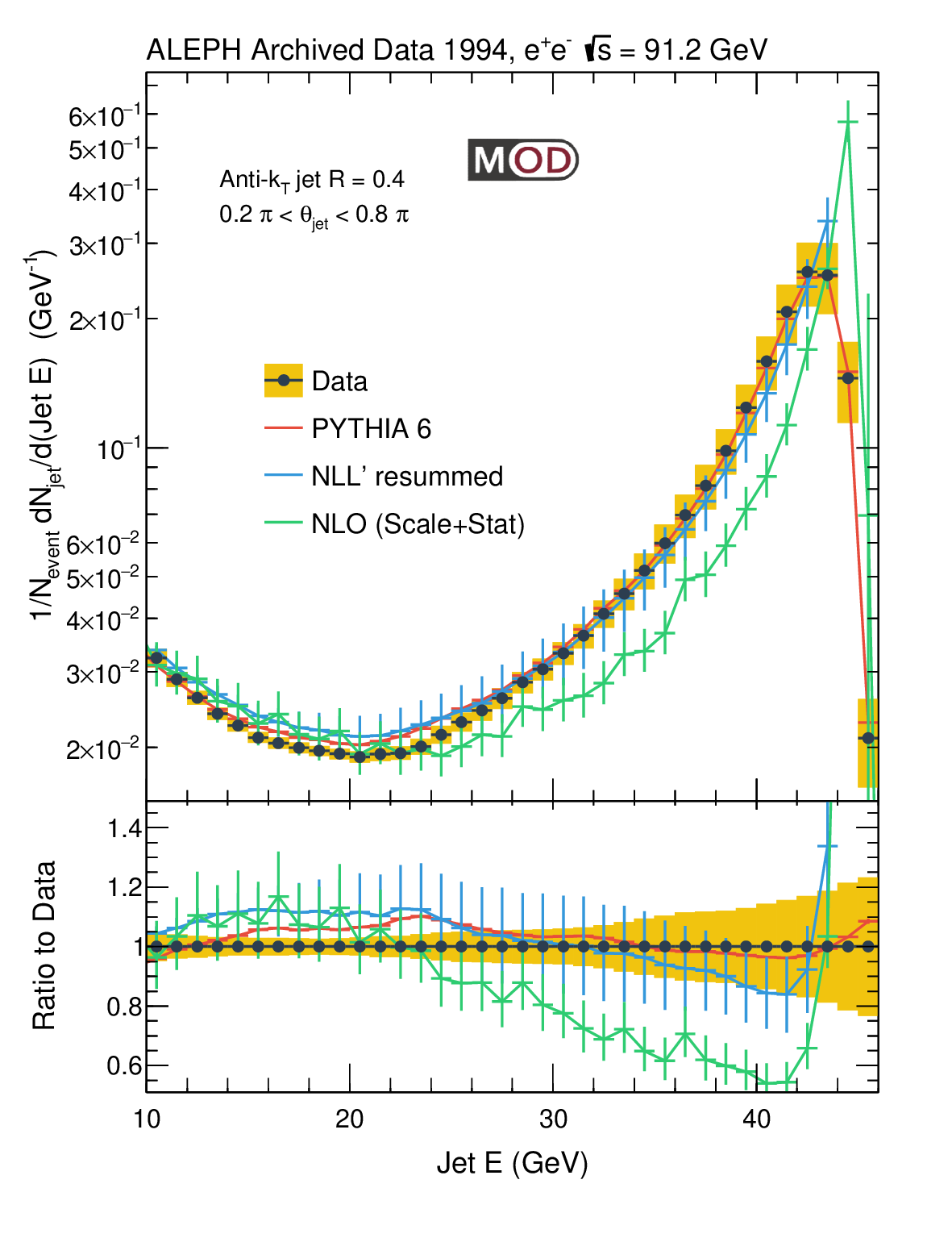}
    \caption{Measured inclusive jet $E$ spectrum.  The spectrum is normalized by the number of events used to perform the measurement.  Any overflow contribution is included in the final bin.  (Left Panel) The data is compared with predictions from \pythia{6} (red), \pythia{8} (blue), \herwig (green), and \sherpa (purple) generators.  The predictions are normalized to have the same area as the data. (Right Panel) The data spectra are compared with pQCD calculations at NLO and NLL'+R resummation~\cite{Neill:2021std}.
    }
    \label{Figure:UnfoldedJetE}
\end{figure}
The first unfolded inclusive jet energy spectrum is shown in Figure~\ref{Figure:UnfoldedJetE}. The distribution is peaked at around half of the $Z^0$ boson mass. This structure is mainly coming from $Z^0\rightarrow q\bar{q}$, with the parton shower of one of the outgoing (anti-)quark is almost fully captured by the anti-$k_T$ algorithm with a resolution parameter of 0.4. 

The spectrum decreases rapidly as one moves to lower jet energy, reaching a minimum at around 20 GeV. Below that, the spectrum increases as we go to even lower energy due to the contributions from soft radiation and combinatorial jets.
These features in the jet energy spectrum are captured by most of the event generators. \pythia{6}, which was tuned to describe ALEPH data, gives the best description of the energy spectrum. At low jet energy, \herwig gives the worse description of the results. It over-predicts the jet spectrum at low jet energy. \pyquen generator, which includes a large jet quenching effect equivalent to that in the inelastic PbPb collisions, predicts a large reduction of the population at around 45 GeV and a significant increase in the number of jets at low jet energy.

The same data could also be compared to a next-to-leading-order perturbative QCD calculation and a calculation based on a next-to-leading logarithmic (NLL’) threshold and jet radius resummation (NLL'+R)~\cite{Neill:2021std}. The NLO calculation predicts a sharper peak at large jet energy. The NLL'+R calculation gives a better description of the inclusive jet energy spectrum, which shows the importance to include the jet radius resummation.

\begin{figure}
    \centering
    \includegraphics[width=0.45\textwidth]{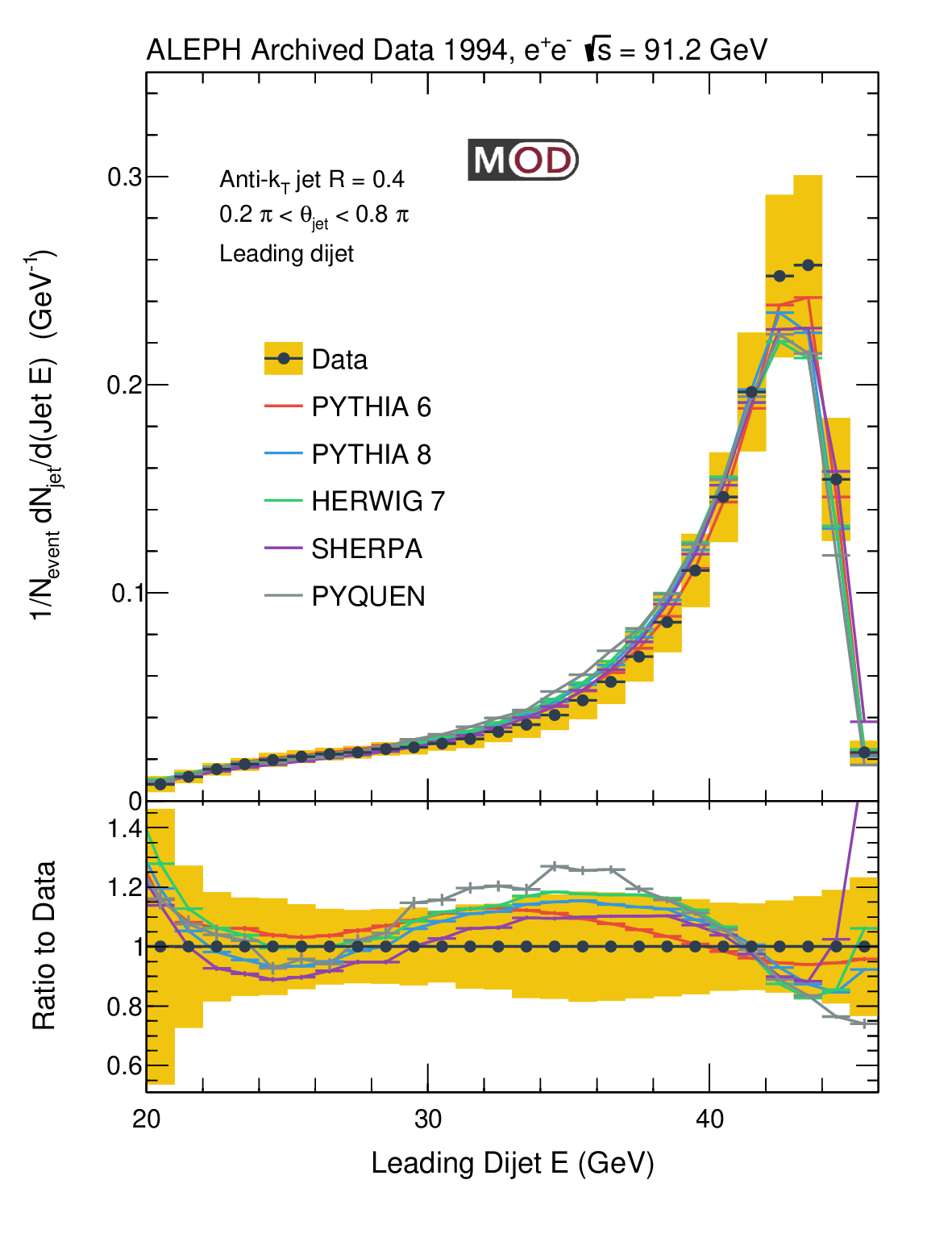}
    \includegraphics[width=0.45\textwidth]{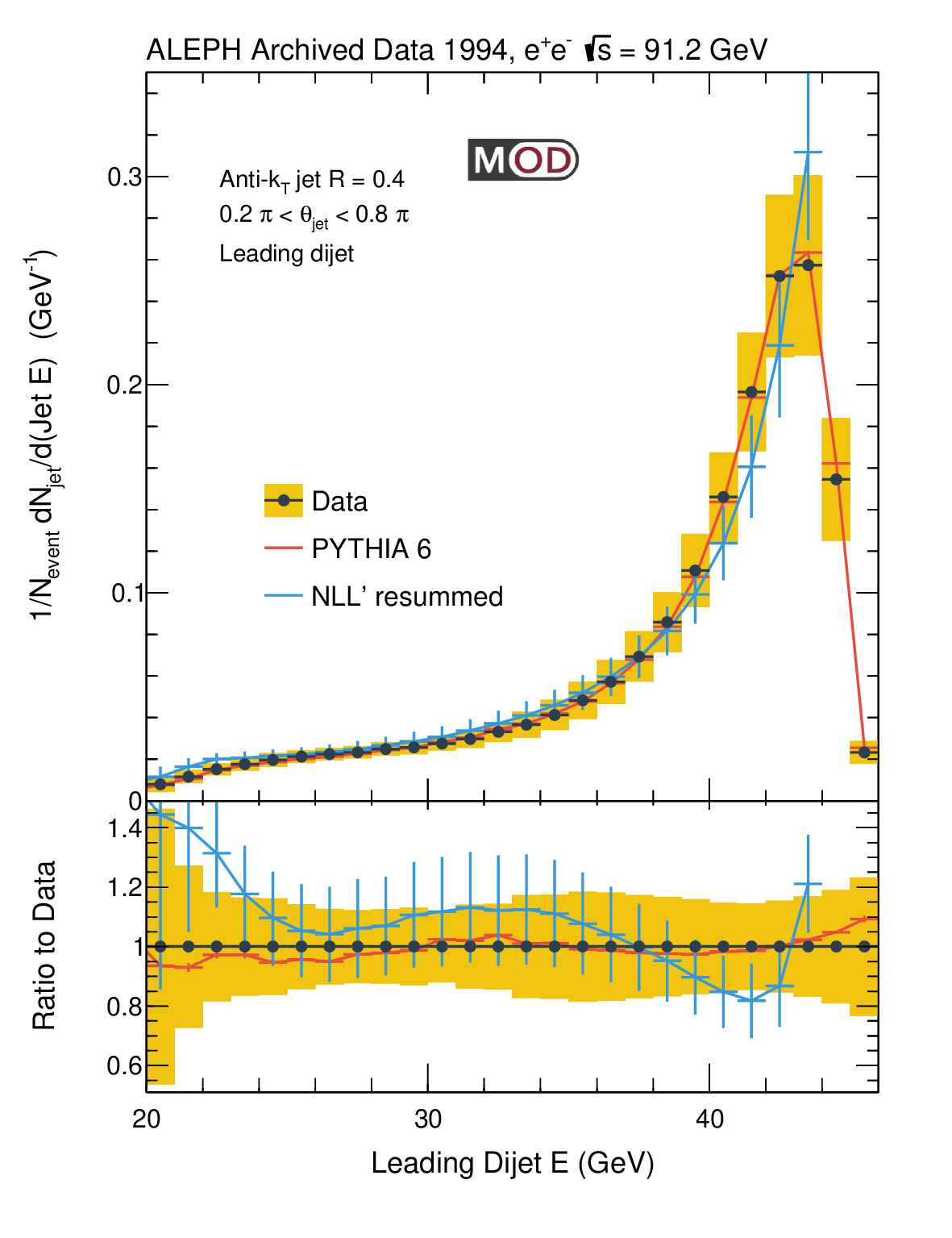}
    \caption{Measured leading dijet jet $E$ spectrum.  The spectrum is normalized by the number of events used to perform the measurement.  Any overflow contribution is included in the final bin.  The data is also compared with predictions from \pythia{6} (red), \pythia{8} (blue), \herwig (green), and \sherpa (purple) generators.  The predictions are normalized to have the same area as the data.  (Right Panel) The data spectra are compared with pQCD calculations at NLL'+R resummation~\cite{Neill:2021std}}
    \label{Figure:UnfoldedJetLeadingDijetE}
\end{figure}

\begin{figure}
    \centering
    \includegraphics[width=0.45\textwidth]{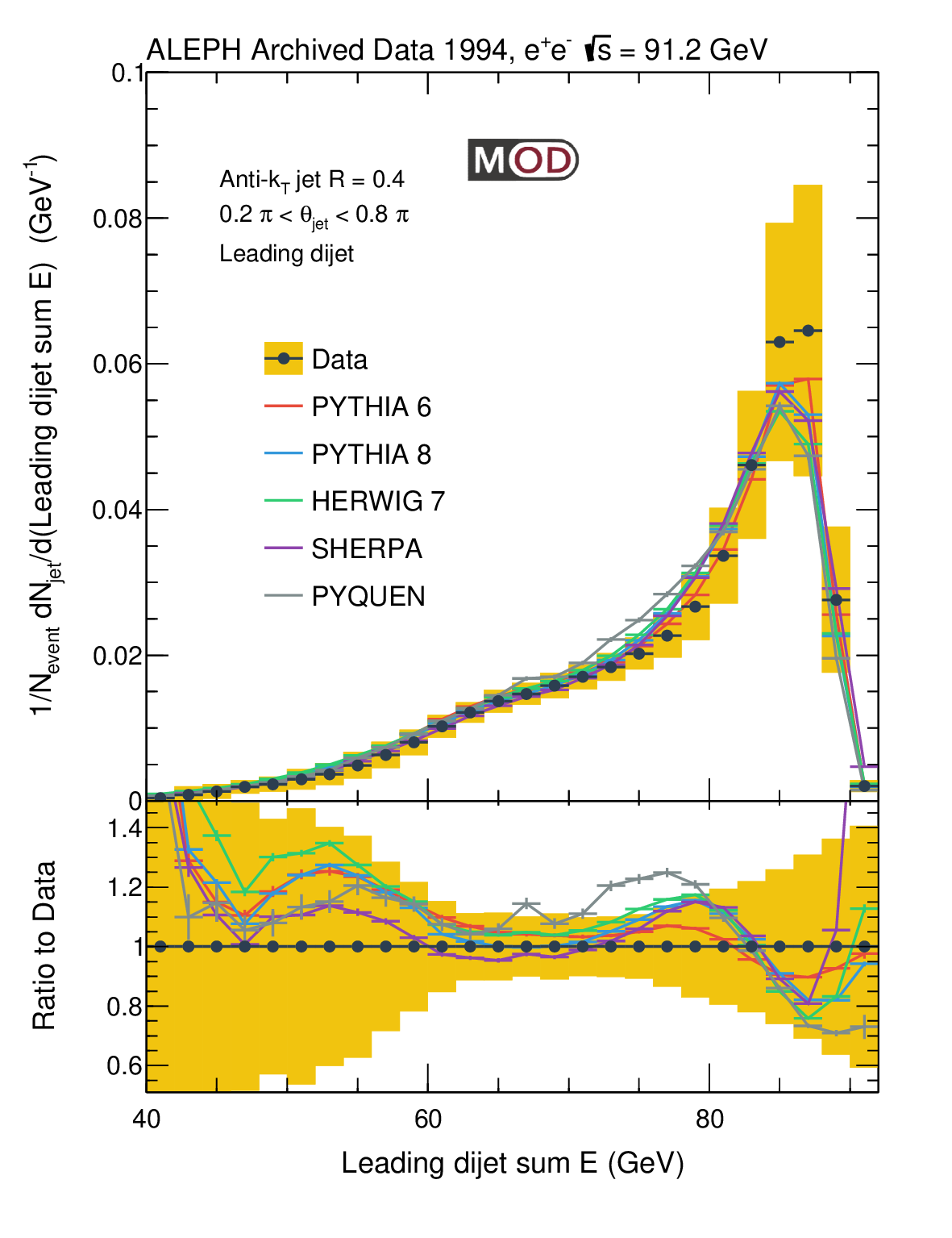}
    \caption{Measured leading dijet sum $E$ spectrum.  The spectrum is normalized by the number of events used to perform the measurement.  Any overflow contribution is included in the final bin.  The data is also compared with predictions from \pythia{6} (red), \pythia{8} (blue), \herwig (green), and \sherpa (purple) generators.  The predictions are normalized to have the same area as the data.}
    \label{Figure:UnfoldedJetLeadingDijetESum}
\end{figure}

In order to focus on the dominant energy flow and suppress the contributions from soft radiation and combinatorial jets, the leading dijet energy is shown in Figure~\ref{Figure:UnfoldedJetLeadingDijetE}. Due to the leading dijet selection, the rise at low jet energy is suppressed. The results are consistent with predictions from \pythia{6} within the quoted systematical uncertainties. \pythia{8}, \herwig, \sherpa tend to overpredict the spectrum at low jet energy. The dijet sum energy (Figure~\ref{Figure:UnfoldedJetLeadingDijetESum}), which is equivalent to the collision energy minus the sum of all the lower energy jets, is more sensitive to the modeling of subleading jets. The levels of (dis)agreement between the simulation and data for leading dijet energy and the leading dijet sum energy are similar.
%\fixme{could consider rearranging figure 13 and 14 to side by side to save some space}

\begin{figure*}
    \centering
    \includegraphics[width=0.9\textwidth]{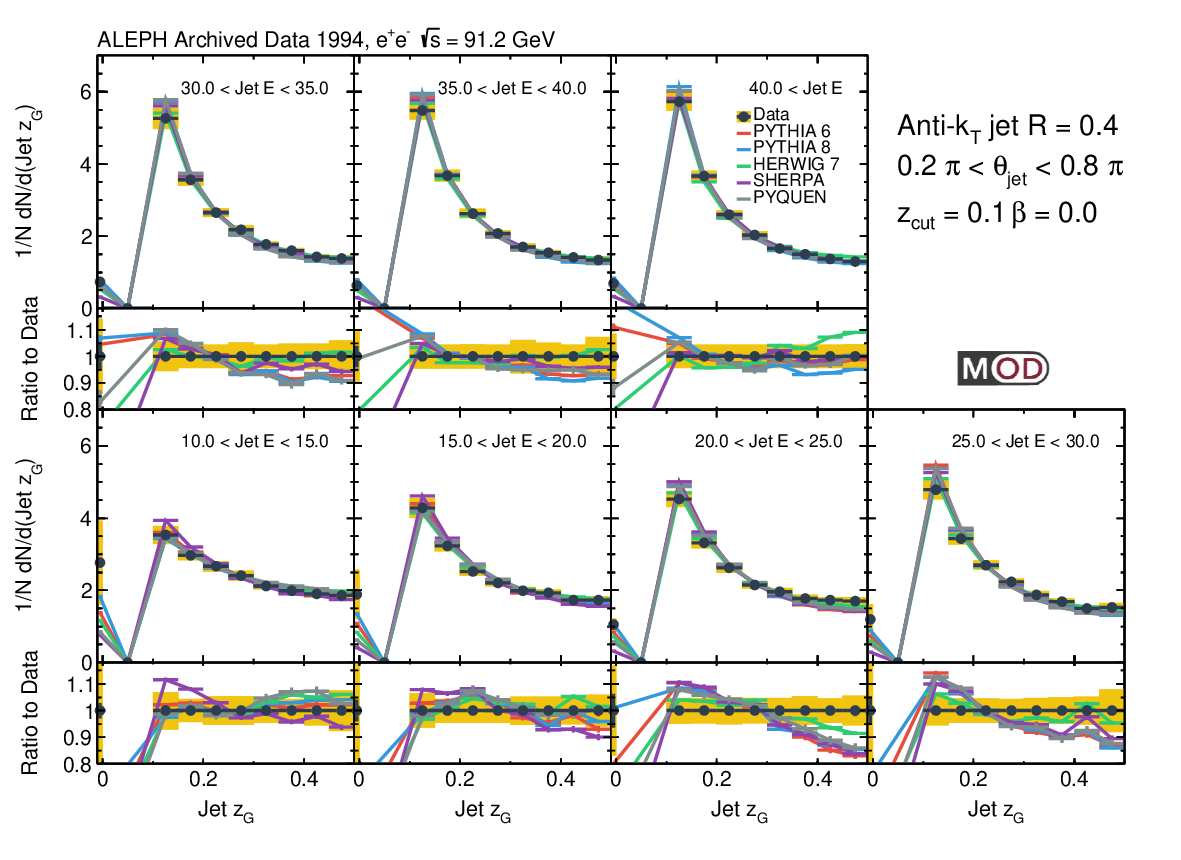}
    \caption{Measured $z_G$ spectra in bins of jet $E$.  The spectrum is self-normalized for each $E$ interval.  The fraction of jets completely groomed away are included in the first bin of each panel.  The data is also compared with predictions from \pythia{6} (red), \pythia{8} (blue), \herwig (green), and \sherpa (purple) generators.}
    \label{Figure:UnfoldedJetZG}
\end{figure*}

\begin{figure*}
    \centering
    \includegraphics[width=0.9\textwidth]{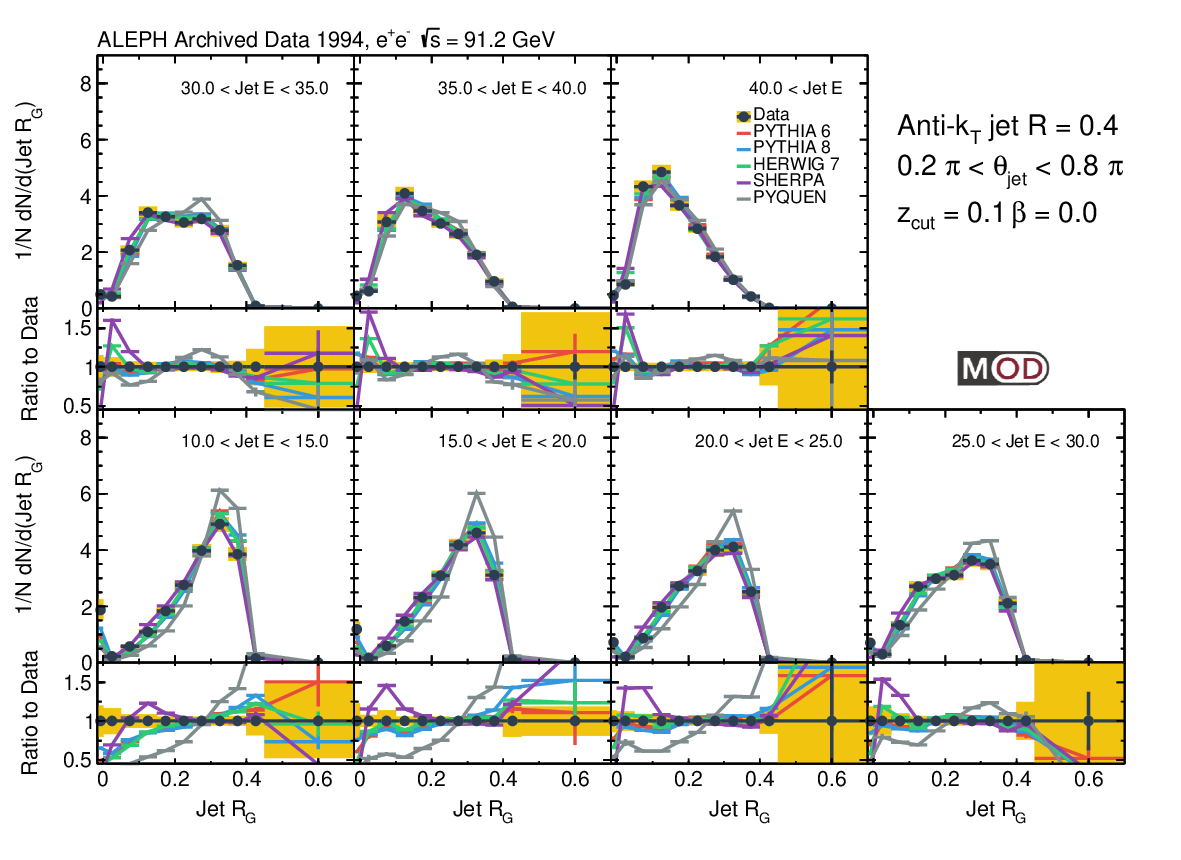}
    \caption{Measured $R_G$ spectra in bins of jet $E$.  The spectrum is self-normalized for each $E$ interval.  The fraction of jets completely groomed away are included in the first bin of each panel.  The data is also compared with predictions from \pythia{6} (red), \pythia{8} (blue), \herwig (green), and \sherpa (purple) generators.}
    \label{Figure:UnfoldedJetRG}
\end{figure*}

Now we change our focus to the characterization of the substructure inside the anti-$k_T$ jets. The groomed momentum sharing $\zg$ spectra and the opening angle of subjets $\Rg$ are presented in bins of jet energy.  The distributions are self-normalized for each energy bin to factor out effects from an imperfect jet energy spectrum in the simulated samples.  As shown in Figure~\ref{Figure:UnfoldedJetZG}, the $\zg$ spectrum is falling as a function of $\zg$ value, reaching a minimum at $\zg\sim 0.5$, which is similar to the data from proton-proton and heavy-ion collisions~\cite{CMS:2017qlm}. The jets which failed the grooming procedure are populated in the first bin $\zg = 0$ in the plot. At high jet energy $E>40$ GeV, \herwig predicts a flatter $\zg$ spectra than data and over-predicts the jets with $\zg$ close to 0.5 while \pythia{6} under-predicts the population at $\zg\sim 0.5$. This is similar to the results from pp collisions at a higher jet transverse momentum~\cite{CMS:2017qlm}. At low jet energy $E<40$ GeV, the data are consistent with predictions from HERWIG 7. In most of the jet energy intervals, \pythia{6}, \pythia{8} and \sherpa under-predict the number of jets at large $\zg$ by up to roughly 10\% while they tend to over-predict the $\zg$ spectra at low $\zg$. 

The $\rg$ is the distance between the groomed subjets. The agreement between $\Rg$ spectra and event generators, as shown in Figure~\ref{Figure:UnfoldedJetRG}, is worse than that observed in the $\zg$ spectra. At high jet energy, event generators predict a slightly narrower core of the $\Rg$ spectra compared to data. At low jet energy, most event generators predict on average a larger separation between subjects (larger $\Rg$) than the unfolded data. The PYQUEN generator that includes jet quenching effect predicts an even larger fraction of jets with large $\Rg$.

\begin{figure*}
    \centering
    \includegraphics[width=0.85\textwidth]{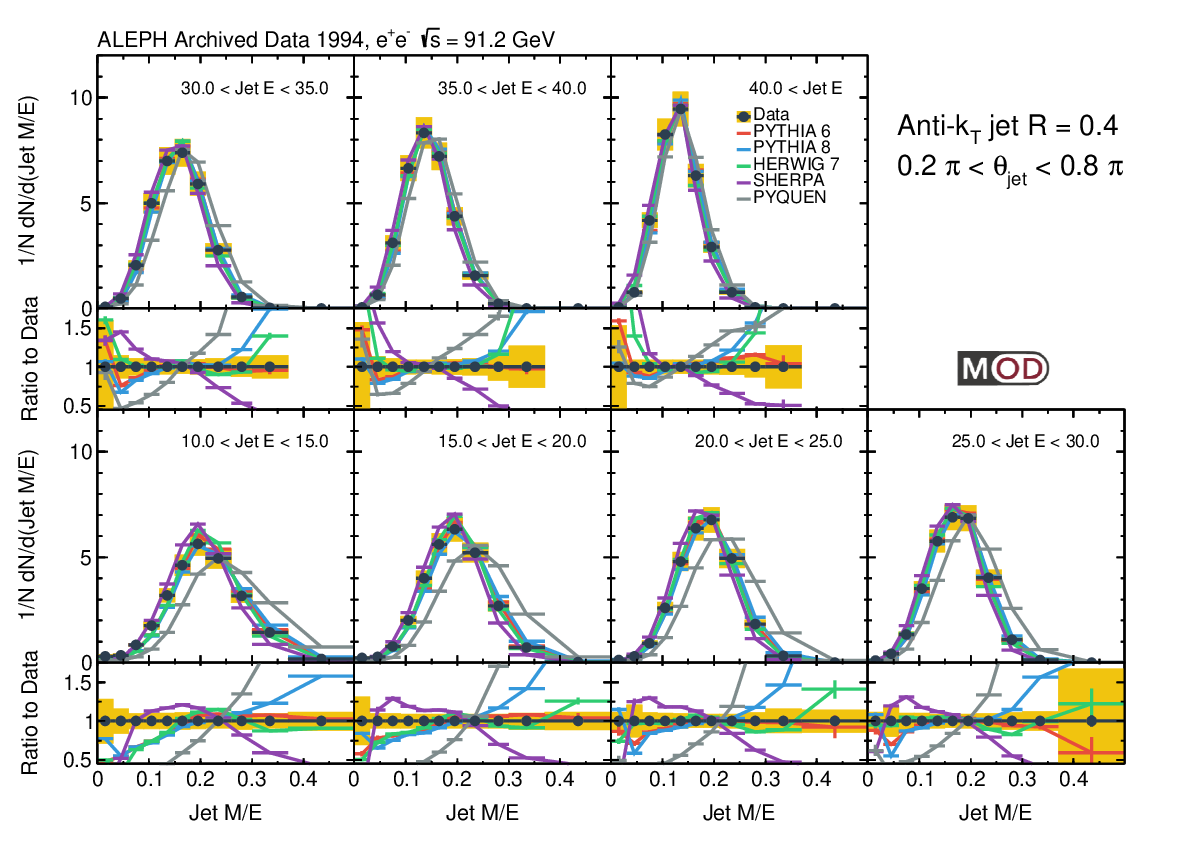}
    \includegraphics[width=0.85\textwidth]{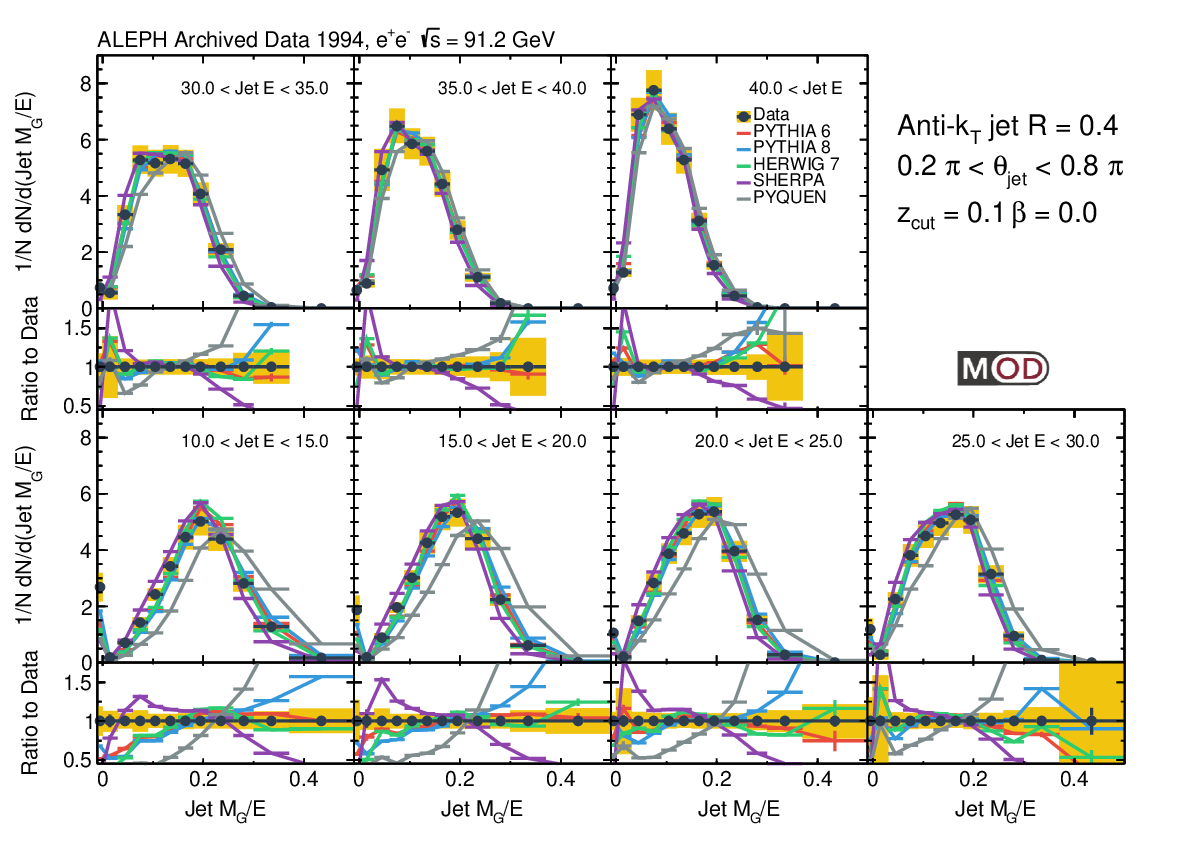}
    \caption{Measured $M/E$ and $M_G/E$ spectra in bins of jet $E$.  The spectrum is self-normalized for each $E$ interval.  The fraction of jets completely groomed away are included in the first bin of each panel.  The data is also compared with predictions from \pythia{6} (red), \pythia{8} (blue), \herwig (green), and \sherpa (purple) generators.}
    \label{Figure:UnfoldedJetMass}
\end{figure*}

The jet mass and groomed jet mass results are shown in Figure~\ref{Figure:UnfoldedJetMass}.  A similar trend to the \Rg result as a function of jet energy is observed: \ME and \MgE are generally smaller at larger jet energy compared to lower energy results.  In addition to the jet transverse width, the mass is also sensitive to the scale of the initial parton.  The difference in agreement between \Rg and mass observables can shed light on the difference in agreement between jet transverse width and the initial scale of the parton.  The \pyquen generator has a more pronounced effect on the mass which manifests as a shift in the measured spectra, which is especially significant at low jet energy.  The difference between \ME and \MgE is also of interest here, and it is sensitive to the amount of wide-angle soft radiation in the jet.  We leave the explicit measurement of $\Delta M/E \equiv (M-M_G)/E$ to future studies.

%%%%%%%%%%%%%%%%%%%%%%%%%%%%%%%%%%%%
\section{\label{Section:Summary}Summary}
%%%%%%%%%%%%%%%%%%%%%%%%%%%%%%%%%%%%
In summary, the first measurements of anti-$k_{T}$ jet energy spectrum and substructure in hadronic $Z^0$ decays are presented. The archived $e^+e^-$ annihilation data at a center-of-mass energy of 91.2 GeV were collected with the ALEPH detector at LEP in 1994. Inclusive jet and dijet energy spectra are presented using jets clustered with a distance parameter $R=0.4$. In addition, jet substructure observables are analyzed as a function of jet energy which includes groomed and ungroomed jet mass to jet energy ratio, groomed momentum sharing, and groomed jet radius. The results are compared with the perturbative QCD calculations and predictions from the \sherpa, \herwig v7.1.5, \pythia{6}, \pythia{8}, and \pyquen event generators in order to test the Monte Carlo modeling of initial state radiation, final state radiation, parton shower, and hadronization effects.
While the inclusive jet and dijet energy spectra agree better with the predictions from the archived \pythia{6} MC, which was tuned to describe the ALEPH data, none of the event generators gives a satisfactory description of the data.
While the inclusive jet spectrum can not be described by parton-level perturbative QCD calculation at next-to-leading order, calculations based on next-to-leading logarithmic threshold and jet radius resummation could give a good description of the leading and inclusive jet energy spectra.
These results provide the cleanest test of the perturbative QCD calculation of jets and serve as an important reference to the studies of jets in proton-proton, heavy-ion, and future electron-ion collisions.

%In summary, the first measurement of two-particle angular correlations for charged particles emitted in $e^+e^-$ collisions at a center-of-mass energy of 91.2 GeV is reported using archived data collected with the ALEPH detector at LEP. The correlation functions are measured over a broad range of pseudorapidity and azimuthal angle of the charged particles. Those results using either lab coordinates or the event thrust coordinates are compared to predictions from the $\textsc{pythia}$, {\sc sherpa} and {\sc herwig} event generators. In contrast to the results from high multiplicity pp, pA and AA collisions, where long-range correlations with large pseudorapidity gap are observed, no significant enhancement of long-range correlations is observed in $e^+e^-$ collisions. The data are compared to generators that do not include additional final-state interactions of the outgoing partons. The results are better described by the {\sc pythia} and {\sc sherpa} generators than {\sc herwig}.  Those results provide new insights to the showering and hadronization modeling and serve as an important reference to the observed long-range correlation in high multiplicity pp, pA and AA collisions.
%%%%%%%%%%%%%%%%%%%%%%%%%%%%%%%%%%%%%%%%%%%%%%%%%%%%%%%%%%%%%%%%%%%%%%%%%%%%%

\begin{acknowledgments}
The authors would like to thank the ALEPH Collaboration for their support and foresight in archiving their data. We thank Joao Pires for providing the NLO calculations of the jet energy spectrum. We appreciate the help from Felix Ringer for providing the calculations with NLL'+R resummation.  We would like to thank the useful comments and suggestions from Roberto Tenchini, Guenther Dissertori, Andrew Lakoski, Liliana Apolinario, Ben Nachman, Camelia Mironov and Jing Wang. 

This work has been supported by the Department of Energy, Office of Science, under Grant No. DE-SC0011088 (to Y.L., Y.C., M.P. and T.S.), Grant No. DE-SC0012567 (to J.T.), Grant No. DE-SC0005131 (to A. Baty), DOE Grants No. DE-SC0018117 and DE-FG02-03ER41244, and the Research Corporation for Scientific Advancement (to D.V.P. and C.M.), the Ministry of Education and the Ministry of Science and Technology of Taiwan (to P.C.), the Harvard Frederick Sheldon Traveling Fellowship (to A. Badea).% \fixme{need update}.
\end{acknowledgments}

%\nocite{*}
\bibliography{ridgepaperALEPH}

\end{document}